\documentclass[a4paper]{article}
\usepackage{graphicx}
\usepackage{amsmath}
\usepackage{multirow}
\usepackage{braket}
\usepackage{cite} 
\usepackage{amsfonts}

\newcommand{\Sd}[1]{S^{\dagger}_{#1}} 
\DeclareMathOperator{\perm}{per}

\usepackage[breaklinks=true,colorlinks=true,linkcolor=blue,urlcolor=blue,citecolor=blue]{hyperref}

\newcommand\undermat[2]{% http://tex.stackexchange.com/a/102468/5764
  \makebox[0pt][l]{$\smash{\underbrace{\phantom{%
    \begin{matrix}#2\end{matrix}}}_{\text{$#1$}}}$}#2}

\begin{document}
\title{Eigenvalue-based determinants for scalar products and form factors in Richardson-Gaudin integrable models coupled to a bosonic mode}

\author{Pieter~W.\ Claeys\textsuperscript{1,2,*} \and  Stijn~De~Baerdemacker\textsuperscript{1,2,3} \and Mario~Van~Raemdonck\textsuperscript{1,2,3} \and Dimitri~Van~Neck\textsuperscript{1,2} \\
{\small \textsuperscript{1} Ghent University, Center for Molecular Modeling, Technologiepark 903, 9052 Ghent, Belgium}\\
{\small\textsuperscript{2} Ghent University, Department of Physics and Astronomy, Proeftuinstraat 86, 9000 Ghent, Belgium}\\
{\small\textsuperscript{3} Ghent University, Department of Inorganic and Physical Chemistry, Krijgslaan 281 (S3), 9000 Ghent, Belgium}  \\
\small \textsuperscript{*} E-mail: pieterw.claeys@ugent.be
}
\date{}
\maketitle

\begin{abstract}
Starting from integrable $su(2)$ (quasi-)spin Richardson-Gaudin XXZ models we derive several properties of integrable spin models coupled to a bosonic mode. We focus on the Dicke-Jaynes-Cummings-Gaudin models and the two-channel $(p+ip)$-wave pairing Hamiltonian. The pseudo-deformation of the underlying $su(2)$ algebra is here introduced as a way to obtain these models in the contraction limit of different Richardson-Gaudin models. This allows for the construction of the full set of conserved charges, the Bethe Ansatz state, and the resulting Richardson-Gaudin equations. For these models an alternative and simpler set of quadratic equations can be found in terms of the eigenvalues of the conserved charges. Furthermore, the recently proposed eigenvalue-based determinant expressions for the overlaps and form factors of local operators are extended to these models, linking the results previously presented for the Dicke-Jaynes-Cummings-Gaudin models with the general results for Richardson-Gaudin XXZ models. 
\end{abstract}

\section{Introduction}
%Inleiding tot Dicke model en p+ip pairing model, verschillende fysische situaties maar onderliggende algemene theorie: integreerbaar. Andere integreerbare systemen: RG systemen. Dicke model=contractielimiet, nu ook antonen voor p+ip pairing model. Connectie kan adiabatisch gebeuren en laat toe om allerlei eigenschappen te onderzoeken van deze modellen. Beschrijving in nieuwe variabelen, laat toe om grootschalige numerieke onderzoeken te doen (ref...). Al gedaan voor Dicke model via ABA en alternatieve pseudovacuum, hier alternatieve manier (renormalisatie) en veralgemening naar pairing model.
%Eerst herhaling van vorige resultaten voor spin RG modellen, daarna pseudodeformatie, dan linken en afleiden determinant-formules.

Integrable models can be used to describe a wide range of physical phenomena, where the exact solvability allows for a numerical treatment beyond the domain of applicability of perturbative and mean-field treatments. One such class of integrable systems is the class of Richardson-Gaudin (RG) systems \cite{gaudin_diagonalisation_1976,richardson_restricted_1963,richardson_exact_1964}. These systems support as many (non-trivial) conserved operators commuting with the Hamiltonian as there are degrees of freedom in the system \cite{dukelsky_colloquium:_2004} and the eigenstates are given by a Bethe Ansatz wavefunction. A set of coupled non-linear equations, the so-called RG or Bethe Ansatz equations, have to be solved in order to determine the variables in the wave function. The number of equations scales linearly with the system size, in contrast to the exponential scaling of the Hilbert space when diagonalizing the Hamiltonian matrix. Another attractive feature of these models is the availability of numerically efficient expressions for overlaps and form factors. Traditionally, these expressions can be obtained by Algebraic Bethe Ansatz methods and result in Slavnov \cite{slavnov_calculation_1989,zhou_superconducting_2002,faribault_exact_2008} or Borchardt/Izergin determinants \cite{borchardt_bestimmung_1857,singer_bijective_2004,izergin_determinant_1992,korepin_quantum_1993}. An alternative approach has been introduced by Faribault \emph{et al.} \cite{faribault_gaudin_2011,el_araby_bethe_2012}, closely related to the Bethe ansatz/ordinary differential equation correspondence \cite{marquette_generalized_2012,guan_heine-stieltjes_2012}, and later generalized by us towards the full class of RG models \cite{claeys_eigenvalue-based_2015}. In this approach an alternative, well-conditioned, set of equations is solved for the eigenvalues of the constants of motion characterizing these integrable models. Subsequently, scalar products and certain form factors can be written as determinants of matrices whose entries only depend on these eigenvalues  \cite{faribault_determinant_2012,claeys_eigenvalue-based_2015}. Interestingly, these expressions can also be seen as partition functions with domain wall boundary conditions \cite{faribault_determinant_2012,tschirhart_algebraic_2014}, for which similar results were obtained in the context of spin chains \cite{izergin_determinant_1992,kostov_inner_2012}.

Initially, the RG models where formulated in terms of $su(2)$-algebras describing spin states or fermion pairs, but it is also possible to introduce a bosonic degree of freedom by means of a limiting procedure \cite{gaudin_diagonalisation_1976,dukelsky_exactly_2004,lerma_h._integrable_2011}. Two main classes of systems can be obtained in this way. Originally, the class of Dicke-Jaynes-Cummings-Gaudin models was obtained from the trigonometric RG model, which includes the Jaynes-Cummings \cite{jaynes_comparison_1963}, the Tavis-Cummings \cite{tavis_exact_1968} and the inhomogeneous Dicke model \cite{dicke_coherence_1954}, all describing the interaction between a (set of) two-level system(s) and a single bosonic electromagnetic mode. Another class of systems considers the coupling of an integrable two-channel $(p+ip)$-wave superfluid to a bosonic mode, which can be obtained as the limiting case of a hyperbolic RG model \cite{lerma_h._integrable_2011}. This model was initially introduced by Dunning \emph{et al.}  \cite{dunning_becbcs_2011} and can be shown to be equivalent to a model which couples Cooper pairs to condensed molecular bosons \cite{hibberd_bethe_2006}. These two classes were later obtained as two distinct cases of integrable Hamiltonians containing a bosonic degree of freedom, starting from a variational approach \cite{birrell_variational_2012}.

The derivation of form factors and overlaps in the eigenvalue-based formalism \cite{faribault_determinant_2012,claeys_eigenvalue-based_2015,tschirhart_algebraic_2014} depends heavily on the existence of a dual state and therefore a (dual) highest weight state in conjunction with a lowest weight. This makes the generalization of results for the $su(2)$-models towards models containing a bosonic degree of freedom far from straightforward, because the $hw(1)$ algebra of a bosonic mode is non compact and therefore lacks a highest weight. Tschirhart and Faribault recently showed how determinant expressions could be found for the form factors of the Dicke-Jaynes-Cummings-Gaudin models by means of the introduction of an intricate alternative Algebraic Bethe Ansatz \cite{tschirhart_algebraic_2014}. In the present paper, we obtain these expressions as a limiting case of a renormalized pseudo-deformed spin model. The pseudo-deformation scheme \cite{de_baerdemacker_richardson-gaudin_2012} was originally proposed as a way to shed light on the singularities arising in the RG equations by connecting the spin models to purely bosonic models. Here this method is used as a way to obtain a bosonic algebra as the contraction limit of a $su(2)$ quasispin algebra. The connection can be made adiabatically in a controlled fashion \cite{claeys_dicke_2015} and allows for an extension of the $su(2)$-based integrable systems towards those containing a bosonic degree of freedom. The purpose of the present paper is twofold. First, we show how the eigenvalue-based formalism for the Dicke model, introduced by Tshirhart and Faribault, can be generalized to the extendend $(p+ip)$ (and related) models in the same way we recently generalized the $su(2)$ XXX results to XXZ models. Second, we present the pseudo-deformation scheme as a unifying framework in which all properties (eigenstates, eigenvalue-based variables, form factors, etc.) of both the Dicke and extended $(p+ip)$ model can be derived in a simple and straightforward way.  

In section 2, the necessary preliminaries for Richardson-Gaudin models will be reviewed, after which the pseudo-deformation scheme is discussed (section 3). The connection with the bosonic models is then made explicit in sections 4 and 5 and determinant expressions are presented for overlaps and normalizations in section 6. It is shown how the results for XXZ RG models can be generalized to these models, further extending the description of Richardson-Gaudin integrable models in terms of 'eigenvalue-based' variables and unifying some previously presented results \cite{tschirhart_algebraic_2014,claeys_eigenvalue-based_2015}.

\section{Richardson-Gaudin models}
\subsection{Definitions}
%Afleiden modellen, Gaudin algebra, voor algemene spin, Hamiltoniaan=lin co
Richardson-Gaudin models are defined by a set of $n$ mutually commuting conserved charges \cite{dukelsky_class_2001,ortiz_exactly-solvable_2005} parametrized as
\begin{equation}\label{rg:com}
R_i=S_i^0+g\sum_{k \neq i}^n \left[\frac{1}{2}X_{ik}(S^{\dagger}_kS_i+S_kS^{\dagger}_i)+Z_{ik}S_i^0S_k^0\right],
\end{equation}
with the set of operators $\{S^{\dagger}_i,S_i,S_i^0\}$ $(i=1\dots n)$ spanning a set of $n$ independent $su(2)_i$ (quasi-)spin algebras
\begin{equation}
[S_i^0,S^\dag_k]=\delta_{ik}S^\dag_k,\qquad[S_i^0,S_k]=-\delta_{ik}S_k,\qquad[S_i^\dag,S_k]=2\delta_{ik}S_k^0,
\end{equation}
with irreps $|s_i,\mu_i\rangle$ associated with each separate algebra $su(2)_i$. These algebras can represent genuine spins, general $(2s_i+1)$-levels by means of a Schwinger representation, or fermion quasispin pairs in a pairing model \cite{talmi_simple_1993}. The introduction of these operators allows for an algebraic formulation of these models, independent of the underlying physical interpretation.

The constraints on the $X$- and $Z$-coefficients for which these operators commute mutually were obtained by Gaudin \cite{gaudin_diagonalisation_1976} and result in a set of equations defining a Gaudin algebra as
\begin{align}\label{rg:algebra}
X_{ij}=-X_{ji}, \qquad Z_{ij}=-Z_{ji},\nonumber\\
X_{ij}X_{jk}-X_{ik}(Z_{ij}+Z_{jk})=0,
\end{align}
which have to hold $\forall i \neq j \neq k =1, \dots,n$. Multiple classes of solutions for these equations have been found \cite{gaudin_diagonalisation_1976,dukelsky_class_2001}, where each class considers $X_{ij}$ and $Z_{ij}$ as odd functions of a set of parameters $\{\epsilon_i\}=\{\epsilon_1,\epsilon_2, \dots, \epsilon_n\}$, such that the Gaudin algebra is defined as $X_{ij}=X(\epsilon_i,\epsilon_j)$ and $Z_{ij}=Z(\epsilon_i,\epsilon_j)$. A model is said to be RG integrable if it has as many conserved charges commuting with the Hamiltonian as degrees of freedom \cite{dukelsky_class_2001,links_algebraic_2003}, so an integrable Hamiltonian can be constructed as a linear combination of the conserved charges (\ref{rg:com}), commuting with all these operators by construction.

\subsection{Diagonalizing integrable Hamiltonians and eigenvalue-based variables}
%particle and hole Bethe ansatz- RG vergelijkingen en eigenwaarden - nieuwe variabelen - correspondentie
Since the conserved charges (\ref{rg:com}) commute mutually, they share a common set of eigenstates. However, since any arbitrary product state can be created by either acting with creation operators on the lowest-weight state or by acting with annihilation operators on the highest-weight state, two separate representations are possible for each eigenstate. These will be discussed simultaneously here. After extending the Gaudin algebra (\ref{rg:algebra}) by adding a set of variables (also called rapidities) $x_{\alpha}$ such that $X_{i\alpha}=X(\epsilon_i,x_{\alpha})$, the following creation/annihilation operators can be defined as 
\begin{equation}
\Sd{\alpha}=\sum_{i=1}^n X_{i\alpha}\Sd{i}, \qquad S_{\alpha}=\sum_{i=1}^n X_{i\alpha}S_i,
\end{equation}
fully determined by a single (possibly complex) variable $x_{\alpha}$, where $X_{i\alpha}=X(\epsilon_i,x_{\alpha})$ extends the Gaudin algebra. Eigenstates of the operators $R_i$ can then be constructed by the repeated action of generalized creation/annihilation operators on an empty/fully-filled vacuum state
\begin{equation}
\ket{\psi}=\left(\prod_{\alpha=1}^N \Sd{\alpha}\right)\ket{\theta}, \qquad \ket{\psi'}=\left(\prod_{\alpha'=1}^{M-N}S_{\alpha'}\right)\ket{\theta'},
\end{equation}
where we have defined the particle-vacuum state $\ket{\theta}=\otimes_{i=1}^n\ket{s_i,-s_i}$ and the hole-vacuum state $\ket{\theta'}=\otimes_{i=1}^n\ket{s_i,s_i}$. The number of excitations $N$ is restricted to $N<M=\sum_i 2s_i$. Both states $\ket{\psi}$ and $\ket{\psi'}$ can represent the same eigenstate of $R_i$, in which case they are referred to as dual states. These states can be shown to be eigenstates if the RG equations
\begin{align}\label{rg:rgeq}
1&+g\sum_{i=1}^n s_i Z_{i\alpha}-g\sum_{\beta \neq \alpha}^N Z_{\beta\alpha}=0 \qquad\text{(particles)}, \nonumber\\
-1&-g\sum_{i=1}^ns_iZ_{i\alpha'}-g\sum_{\beta' \neq \alpha'}^{M-N}Z_{\beta'\alpha'}=0 \qquad\text{(holes)},
\end{align}
are satisfied. The resulting eigenvalues $r_i$ of the conserved charges $R_i$ are then given by
\begin{align}\label{rg:ri}
r_i&=d_i\left(-1-g\sum_{\alpha=1}^N Z_{i \alpha}+g\sum_{k \neq i}^n Z_{ik}d_k\right)  \qquad\text{(particles)},\nonumber\\
r_i'&=d_i\left(1-g\sum_{\alpha'=1}^{M-N} Z_{i\alpha'}+g\sum_{k \neq i}^n Z_{ik} d_k\right) \qquad\text{(holes)}.
\end{align}

Following on recent work on the XXX RG model \cite{faribault_gaudin_2011,el_araby_bethe_2012}, we showed \cite{claeys_eigenvalue-based_2015} how an alternative description of the Bethe Ansatz states could be obtained starting from a new set of variables
\begin{equation}
\Lambda_i=\sum_{\alpha=1}^N Z_{i\alpha}, \qquad \Lambda_i'=\sum_{\alpha'=1}^{M-N}Z_{i\alpha'}, \qquad \forall i=1 \dots n.
\end{equation}
It can be seen that each eigenvalue $r_i$ (\ref{rg:ri}) is only explicitly dependent on the parameter $\Lambda_i$ (particle) or $\Lambda_i'$ (hole). This has led to the denomination 'eigenvalue-based variables'. Since the eigenvalue of an eigenstate is independent of its particular representation, a correspondence between $\Lambda_i$ and $\Lambda_i'$ can be found by equating the eigenvalues $r_i=r_i'$, leading to
\begin{equation}\label{rg:condualvar}
g\Lambda_i'=g\Lambda_i+2,  \qquad \forall i=1 \dots n.
\end{equation}
So, if the eigenvalue-based variables are known in one representation, those for the dual representation immediately follow. Remarkably, the overlaps with non-interacting states (such as single Slater determinants for the quasispin picture), as well as the normalization and several form factors can also be determined solely from these variables (see the next subsection).

\subsection{Determinant expressions}
%overlap met Slater determinant (algemeen=permanent, voor spin-1/2 determinant) - twee mogelijkheden - twee manieren om elke toestand te schrijven - overlap van BA met pseudovacuum - laat toe om vorm factoren te geven
When performing calculations with Bethe Ansatz states, it is often convenient to expand them in a basis set. The wavefunction can be expanded in the complete set of basis states
\begin{equation}\label{de:sd}
\ket{\{N_i\}}=\prod_{i=1}^n \left(S^{\dagger}_i\right)^{N_i}\ket{\theta},
\end{equation}
with the integers $[N_1,\dots,N_n]\equiv [\{N_i\}]$ a partitioning of the number of excitations $N$ over the number of levels $n$. The expansion is given by 
\begin{equation}\label{det:expansion}
\prod_{\alpha=1}^N\left(\sum_{i=1}^n X_{i\alpha}S^{\dagger}_i\right)\ket{\theta}=\sum_{[\{N_i\}]}\phi_{[\{N_i\}]}\prod_{i=1}^n \left(S^{\dagger}_i\right)^{N_i}\ket{\theta},
\end{equation}
with expansion coefficients given by the permanent of a matrix \cite{percus_combinatorial_1971,johnson_size-consistent_2013}
\begin{equation}\label{det:expansioncoeff}
\phi_{[\{N_i\}]}=\frac{1}{N_1! \dots N_n!} \perm \left(C^N_{[\{N_i\}]}\right)
\end{equation}
and
\begin{equation}
C^N_{[\{N_i\}]}=\left(
\begin{array}{ccccccc}
X_{i_1 \alpha_1} & \dots & X_{i_1 \alpha_1} & \dots & X_{i_N \alpha_1} & \dots & X_{i_N \alpha_1}\\
X_{i_1 \alpha_2} & \dots & X_{i_1 \alpha_2} &  \dots & X_{i_N \alpha_2} & \dots & X_{i_N \alpha_2} \\
\vdots & & \vdots & & \vdots & & \vdots\\
\undermat{N_1}{X_{i_1 \alpha_N} & \dots & X_{i_1 \alpha_N}} &  \dots & \undermat{N_N}{X_{i_N \alpha_N} & \dots & X_{i_N \alpha_N}}
\end{array}\right)
.
\end{equation}
\vspace{\baselineskip}

\noindent For doubly-degenerate models ($s_i=1/2, \forall i$), the occupation $N_i$ of each level is either $0$ or $1$ and this permanent can be rewritten as a determinant \cite{claeys_eigenvalue-based_2015}, which can efficiently be evaluated numerically. Labelling the set of occupied levels ($N_i=1$) as $\{i_a\}=\{i_1, \dots , i_N\}$, the overlap of a Bethe Ansatz state with
\begin{equation}
\ket{\{i_a\}}=\ket{\{i_1\dots i_N\}}=\left(\prod_{a=1}^N S^{\dagger}_{i_a}\right)\ket{\downarrow \dots \downarrow}
\end{equation}
is given by 
\begin{equation}\label{ov:XXZ}
\phi_{[\{N_i\}]}=\braket{\{i_1\dots i_N\}|\{x_{\alpha}\}}=\frac{\prod_{\alpha=1}^NX_{r\alpha}}{\prod_{a=1}^NX_{r i_a}} \det J
\end{equation}
with
\begin{equation}
J_{ab}=\begin{cases}
\Lambda_{i_a}-\sum_{c \neq a}^NZ_{i_ai_c}+Z_{ri_a}  &\text{if}\ a=b\\
X_{i_ai_b}  &\text{if}\ a \neq b\\
\end{cases}.
\end{equation}
Here the Gaudin algebra (\ref{rg:algebra}) was extended by means of an arbitrary gauge variable $\epsilon_r$, as introduced in \cite{claeys_eigenvalue-based_2015}. It is worth noting that the overlaps $\braket{\{i_1, \dots, i_N\} | \{x_{\alpha}\}}$ (Eq. (\ref{ov:XXZ})) are implicitly independent of the gauge variable $\epsilon_r$, so the freedom left in the choice of this variable can be exploited to obtain the most simple expression possible for each realization of the Gaudin algebra, usually by setting $\epsilon_r=0$ or $\infty$. This expression for the overlap, combined with the existence of the dual representation, are the key ingredients for the following results. The overlap between a state and its dual state is given by 
\begin{equation}
\braket{\{x_{\alpha}'\}|\{x_{\alpha}\}}=\braket{\theta'|\left(\prod_{\alpha'=1}^{M-N}\Sd{\alpha'}\right)\left(\prod_{\alpha=1}^{N}\Sd{\alpha}\right)|\theta},
\end{equation}
which can also be seen as the overlap of a Bethe Ansatz state with $M$ excitations defined by the set of variables $\{x_{\alpha}\}\cup\{x_{\alpha'}\}$ with the hole-vacuum
\begin{equation}\label{ov:norm}
\braket{\{x_{\alpha}'\}|\{x_{\alpha}\}}=\braket{\theta'|\prod_{\mu=\alpha, \alpha'}^M\Sd{\mu}|\theta}.
\end{equation}
For this state, the eigenvalue-based variables appearing in the diagonal elements of the overlap matrix are given by
\begin{equation}
\Lambda_i^{tot}=\sum_{\alpha=1}^N Z_{i\alpha}+\sum_{\alpha'=1}^{M-N}Z_{i\alpha'}=\Lambda_i+\Lambda_i'=2\Lambda_i+\frac{2}{g},
\end{equation}
where we have used the relation (\ref{rg:condualvar}) between these variables for the dual representation. The overlap (\ref{ov:norm}) can then be written as
\begin{equation}
\braket{\{x_{\alpha}'\}|\{x_{\alpha}\}}=\frac{\prod_{\alpha}X_{r\alpha}}{\prod_i X_{ri}} \det J,
\end{equation}
with
\begin{equation}
J_{ij}=\begin{cases}
2\Lambda_{i}+\frac{2}{g}-\sum_{k \neq i}^n Z_{ik}+Z_{ri}  &\text{if}\ i=j \\
X_{ij}  &\text{if}\ i \neq j\\
\end{cases}.
\end{equation}
Note that this does not depend explicitly on the dual state, but only on the original state through the terms $\Lambda_i$ in the diagonal elements. Once this overlap is known, together with the overlap of both representations with an arbitrary basis state, all information is present to calculate the normalization of both states, as originally shown for the RG XXX model \cite{faribault_determinant_2012}.

\section{Pseudo-deformation of the quasispin}
%Contractie - manier van bosonisatie - alternatief voor Holstein-Primakoff - Dukelsky (atoom-molecule) - link BCS/BA met condensaat van bosonen via adiabatische connectie
%algebra definieren, tonen hoe het kan gebruikt worden als RG solver voor algemene RG systemen
The results reviewed in the previous section can be generalized to integrable systems containing a bosonic degree of freedom by replacing one of the $su(2)$ spin algebras with a bosonic $hw(1)$ algebra. There are multiple approaches possible for this process of bosonization, of which the Holstein-Primakoff transformation \cite{holstein_field_1940} is arguably the most well-spread. Here we have opted to use the recently proposed pseudo-deformation scheme \cite{de_baerdemacker_richardson-gaudin_2012} as a way to obtain bosonic commutation relations because this method provides an adiabatic, and therefore controlled, mapping of the hard-core bosonic $su(2)$ algebra on a genuinely bosonic algebra.

A pseudo-deformed $su(2)_{\xi}$ algebra can be defined as
\begin{equation}\label{pseudo:def}
[S^0(\xi),S^\dag(\xi)]=S^\dag(\xi),\quad[S^0(\xi),S(\xi)]=-S(\xi),\quad[S^\dag(\xi),S(\xi)]=2\left(\xi S^0(\xi) +(\xi-1)s\right)
\end{equation}
with $\xi \in [0,1]$ the pseudo-deformation parameter and $s$ the original ($\xi=1$) $su(2)$ irrep label. This definition can be interpreted as providing a linear interpolation between two known limits: $\xi=1$ gives rise to the original $su(2)$ quasispin algebra, while $\xi=0$ results in a (unnormalized) bosonic $hw(1)$ algebra. This latter limit was also termed the contraction limit of the algebra \cite{gilmore_lie_2008}. The nomenclature \emph{pseudo}-deformation was originally proposed because this algebra can be reduced to a canonical $su(2)_{\xi}$ algebra
\begin{equation}\label{pseudo:su2}
[A^0(\xi),A^\dag(\xi)]=A^\dag(\xi),\qquad[A^0(\xi),A(\xi)]=-A(\xi),\qquad[A^\dag(\xi),A(\xi)]=2A^0(\xi),
\end{equation}
by defining
\begin{equation}
A^\dag(\xi)=\frac{1}{\sqrt{\xi}}S^\dag(\xi),\quad A(\xi)=\frac{1}{\sqrt{\xi}}S(\xi),\quad A^0(\xi)=S^0(\xi)+\left(1-\frac{1}{\xi}\right)s,
\end{equation}
except for the contraction limit $\xi=0$. In this limit, the following operators
\begin{equation}
b^\dag=\sqrt{\frac{1}{2s}}S^\dag(0),\quad b=\sqrt{\frac{1}{2s}}S(0),\quad b^\dag b=S^0(0)+s,
\end{equation}
close the bosonic $hw(1)$ algebra
\begin{equation}
[b^\dag b,b^\dag]=b^\dag,\quad [b^\dag b,b]=-b,\quad [b,b^\dag]=1.
\end{equation}
The irreducible representations of the $\{A^\dag(\xi),A(\xi),A^0(\xi)\}$ algebra are labeled by $s(\xi)\equiv s/\xi$. The interpretation behind this is a gradual increase of the effective multiplicity $(2s(\xi)+1)$ of the $su(2)_{\xi}$ irrep with decreasing $\xi$. It should be noted that only discrete values of $\xi_n=\frac{2s}{n}$ (with $n=2s,2s+1,\dots$) give rise to unitary irreps.  Nevertheless, this is not problematic because the theory of RG integrability is not based on matrix representations (with integer dimensions), so the parameter $\xi$ can be regarded as a continuous variable.

This construction has led to a numerical solution method for the RG equations (\ref{rg:rgeq}) by solving the equations adiabatically from the contraction limit ($\xi=0$) to the $\xi=1$ case \cite{de_baerdemacker_richardson-gaudin_2012,van_raemdonck_exact_2014}. Because the $\{A^0(\xi),A^{\dagger}(\xi),A(\xi)\}$ operators span a canonical $su(2)_{\xi}$ algebra, the set
\begin{equation}
R_i(\xi)=A_i^0(\xi)+g \xi \sum_{k \neq i}^n \left[\frac{1}{2}X_{ik}(A^{\dagger}_k(\xi)A_i(\xi)+A_k(\xi)A^{\dagger}_i(\xi))+Z_{ik}A_i^0(\xi)A_k^0(\xi)\right], \qquad \forall i=1, \dots ,n,
\end{equation}
remains in involution for every value of $\xi \neq 0$ provided the Gaudin equations (\ref{rg:algebra}) are satisfied. The Bethe Ansatz state in the particle representation for these operators is then given by
\begin{equation}
\ket{\psi}=\prod_{\alpha=1}^N\left(\sum_{i=1}^n X_{i\alpha}A^{\dagger}_i(\xi) \right)\ket{\theta},
\end{equation}
if the pseudo-deformed RG equations
\begin{equation}
1+g \sum_{i=1}^n Z_{i\alpha}s_i -g \xi \sum_{\beta \neq \alpha}^N Z_{\beta \alpha}=0, \qquad \forall \alpha=1 \dots N,
\end{equation}
are satisfied. The key observation here is that the set of coupled non-linear RG equations (\ref{rg:rgeq}) reduce to a single uncoupled equation in the contraction limit $\xi=0$, formally equivalent to the secular equation of the $pp$-TDA \cite{de_baerdemacker_tamm-dancoff_2011}. This equation can be straightforwardly solved numerically, after which these solutions can be adiabatically connected to the RG equations of the original problem (\ref{rg:rgeq}) by slowly tuning $\xi \to 1$, as proposed and illustrated in \cite{de_baerdemacker_richardson-gaudin_2012,van_raemdonck_exact_2014}.

\section{Integrable models containing a bosonic degree of freedom}
%inleiding, beide Hamiltonianen geven met wat uitleg erbij. Een algebra uitkiezen uit een set van n+1 en van deze de contractielimiet nemen. Singuliere limiet, maar als we een parametrizatie vinden waarvoor deze regulier is in de limiet bekomen we bosonische commutatierelaties en blijft de volledige RG constructie geldig.
In the previous section a fully-bosonic integrable model was obtained by deforming all $su(2)$ algebras simultaneously into $hw(1)$. A natural digression would be to consider a situation where only one of the spin algebras is pseudo-deformed, leading to an interacting boson in the contraction limit.  Since the bosonic limit is a singular limit of the algebra, care has to be taken to obtain finite results in the contraction limit. However, if a model is found for which no singularities arise in this limit, an RG integrable model is obtained. Several integrable models are known which also contain an interacting boson, of which the Dicke-Jaynes-Cummings-Gaudin model (henceforth referred to as the Dicke model), is the best-known. We will detail the derivation for the Dicke model, and summarize the key results for the extended $(p+ip)$ model, due to the large similarity with the Dicke model.

\subsection{The Dicke model}
%Afleiden, link met Dukelsky vermelden,Hamiltoniaan+BA+RG vergelijkingen
The Dicke Hamiltonian \cite{dicke_coherence_1954} is given by
\begin{equation}
H=\epsilon_0 b^{\dagger}b+\sum_{i=1}^m \epsilon_i S^0_i +g \sum_{i=1}^m \left(\Sd{i} b +S_i b^{\dagger}\right)
\end{equation}
and describes a set of $m$ two-level systems ($s_i=1/2$) interacting with a single mode of the bosonic field, represented by a photon with energy $\epsilon_0$. The connection between this model and the XXZ RG systems was already made by Gaudin \cite{gaudin_diagonalisation_1976} and later extended by Dukelsky \emph{et al.} \cite{dukelsky_exactly_2004}. The derivation presented here is similar to the one by Dukelsky \emph{et al.}, but differs in our choice of bosonization scheme \cite{dukelsky_class_2001}. Starting from the constants of motion for a set of $m+1$ spin systems, we label them $i=0,1, \dots, m$ and exchange the $su(2)$ algebra labeled $i=0$ by a pseudo-deformed algebra $su(2)_{\xi}$. Starting from the trigonometric Gaudin algebra \cite{ortiz_exactly-solvable_2005}
\begin{equation}
X_{ij}=\frac{\sqrt{(1+\eta_i^2)(1+\eta_j^2)}}{\eta_i-\eta_j}, \qquad Z_{ij}=\frac{1+\eta_i \eta_j}{\eta_i-\eta_j}
\end{equation}
and taking the limit $\eta_0 \to \infty$, a Gaudin algebra is obtained determined by
\begin{align}
X_{0k}&=\sqrt{1+\eta_k^2}, \qquad Z_{0k}=\eta_k, \\
X_{ik}&=\frac{\sqrt{(1+\eta_i^2)(1+\eta_j^2)}}{\eta_i-\eta_j}, \qquad Z_{ik}=\frac{1+\eta_i \eta_j}{\eta_i-\eta_j}.
\end{align}
The conserved charge associated with the deformed algebra is given by 
\begin{align}
R_0(\xi)&=A^0(\xi)+g\sum_{k \neq 0}^m \left[\frac{1}{2}X_{0k}\left(A^{\dagger}(\xi)S_k+A(\xi)S^{\dagger}_k\right)+Z_{0k} A^0(\xi)S_i^0\right]\nonumber\\
&=A^0(\xi)+g\sum_{k \neq 0}^m \left[\frac{1}{2}\sqrt{1+\eta_k^2}\left(A^{\dagger}(\xi)S_k+A(\xi)S^{\dagger}_k\right)\eta_k A^0(\xi)S_i^0\right]. 
\end{align}
It is now possible to define a $\xi$-dependent coupling constant as $g=\sqrt{\frac{2\xi}{s_0G^2}}\frac{G^2}{\epsilon_0}$ and rescale the variables $\eta_k=-\sqrt{\frac{\xi}{2s_0G^2}}\epsilon_k$. Near the contraction limit ($\xi \approx 0$) the Gaudin algebra now reduces to
\begin{equation}\label{dicke:param}
X_{0k}=1+\frac{\xi}{4s_0 G^2}\epsilon_k^2+\mathcal{O}(\xi^2), \qquad Z_{0k}=-\sqrt{\frac{\xi}{2s_0G^2}}\epsilon_k,
\end{equation}
which is related to the parametrization proposed by Dukelsky \emph{et al.} \cite{dukelsky_exactly_2004}. By making use of this parametrization, the Dicke Hamiltonian can be obtained (up to a diverging constant) as
\begin{equation}
R_0^D \equiv \epsilon_0R_0(\xi \to 0)=\epsilon_0b^{\dagger}b+\sum_{i=1}^m \epsilon_iS^0_i +g \sum_{i=1}^m \left(\Sd{i} b +S_i b^{\dagger}\right)
\end{equation}
in the contraction limit. A similar procedure leads to
\begin{enumerate}
\item the other constants of motion $R_i^D$ \cite{dukelsky_class_2001}
\begin{equation}\label{dicke:com}
R_i^D \equiv \epsilon_0 R_i(\xi \to 0)=(\epsilon_0-\epsilon_i)S^0_i-G(\Sd{i}b+S_i b^{\dagger})-2G^2\sum_{k \neq i}^m \frac{1}{\epsilon_i-\epsilon_k}\left[\frac{1}{2}(\Sd{i}S_k+S_i \Sd{k})+S^0_iS^0_k\right],
\end{equation}
\item the Bethe Ansatz for the Dicke model \cite{tsyplyatyev_simplified_2010}
\begin{equation}\label{dicke:ba}
\ket{\psi}=\left(\frac{2s_0}{\xi}\right)^{\frac{N}{2}}\prod_{\alpha=1}^N\left(b^{\dagger}-G\sum_{i=1}^m \frac{\Sd{i}}{\epsilon_i-x_{\alpha}}\right)\ket{\theta},
\end{equation}
where the prefactor can be absorbed in the normalization, and

\item the RG equations \cite{tsyplyatyev_simplified_2010}
\begin{equation}
(\epsilon_0-x_{\alpha})-2G^2 \sum_{i=1}^m \frac{s_k}{\epsilon_k-x_{\alpha}}+2G^2 \sum_{\beta \neq \alpha}^N \frac{1}{x_{\beta}-x_{\alpha}}=0, \qquad \forall \alpha=1 \dots N.
\end{equation}
\end{enumerate}
Remarkably, it is also possible to transform the equations for the rapidities to an equivalent set of equations for the set of eigenvalue-based variables provided $s_i=1/2, \forall i=1,\dots,m$ \cite{babelon_bethe_2007,faribault_gaudin_2011}. For the Dicke model, these variables are given by
\begin{equation}
\Lambda_i=\sum_{\alpha=1}^N \frac{1}{\epsilon_i-x_{\alpha}}, \qquad i=1 \dots m.
\end{equation}
These determine the eigenvalues of the constants of motion (\ref{dicke:com}) as 
\begin{align*}
R_i^D \ket{\psi}&=\frac{1}{2}\left[(\epsilon_i-\epsilon_0)+2G^2\sum_{\alpha=1}^N \frac{1}{\epsilon_i-x_{\alpha}}-G^2 \sum_{k \neq i}^m \frac{1}{\epsilon_i-\epsilon_j}\right]\ket{\psi} \\
&=\frac{1}{2}\left[(\epsilon_i-\epsilon_0)+2G^2\Lambda_i-G^2 \sum_{k \neq i}^m \frac{1}{\epsilon_i-\epsilon_j}\right]\ket{\psi}, 
\end{align*}
and satisfy the coupled quadratic equations
\begin{equation}
G^2\Lambda_i^2=N-\Lambda_i(\epsilon_i-\epsilon_0)+G^2\sum_{j\neq i}^m\frac{\Lambda_i-\Lambda_j}{\epsilon_i-\epsilon_j}, \qquad \forall i=1 \dots n.
\end{equation}
These equations can either be determined starting from the RG equations for the Dicke model \cite{babelon_bethe_2007,faribault_gaudin_2011,tschirhart_algebraic_2014}, or by taking the contraction limit of the eigenvalue-based equations for the trigonometric RG model (see Appendix \ref{app:ebv}). The quadratic equations can easily be solved numerically, whereas the RG equations become singular near the so-called singular points \cite{richardson_numerical_1966,dominguez_solving_2006}. Similar to the $su(2)$-based RG models, it is possible to express the overlap and form factors of the Dicke model in the eigenvalue-based variables only, circumventing the need to calculate the singularity-prone rapidities \cite{tschirhart_algebraic_2014} (see also section 6). 

\subsection{The $(p+ip)$-wave pairing model}
\label{pip}
%Afleiden, link met fase diagram p+ip (Rombouts)
A similar procedure can be used to prove the RG integrability of the $(p+ip)$-wave pairing model coupled to a bosonic degree of freedom. This model was introduced by Dunning \emph{et al.} \cite{dunning_becbcs_2011} as an extension of the integrable fermionic $p_x+ip_y$-pairing model \cite{ibanez_exactly_2009,rombouts_quantum_2010,dunning_exact_2010}. Lerma \emph{et al.} consequently showed how this model is given by the limit of a hyperbolic RG model \cite{lerma_h._integrable_2011}, which we will reformulate by making use of the pseudo-deformation. In addition, we will show how the model fits within the eigenvalue-based language. The Hamiltonian is given by 
\begin{eqnarray}
\hat{H}&=&\delta\, b^{\dagger}b+\sum_{\mathbf{k}}\frac{\mathbf{k}^2}{2m}c^{\dagger}_{\mathbf{k}}c_\mathbf{k}-\frac{G}{4}(k_x-ik_y)(k_x'+ik_y')\sum_{\mathbf{k} \neq \pm \mathbf{\mathbf{k}}'}c^{\dagger}_{\mathbf{k}}c^{\dagger}_{-\mathbf{k}}c_{\mathbf{k}'}c_{-\mathbf{k}'} \nonumber \\
&&-\frac{K}{2}\sum_{\mathbf{k}}\left((k_x-ik_y)c^{\dagger}_{\mathbf{k}}c^{\dagger}_{-\mathbf{k}}b+h.c.\right),
\end{eqnarray}
and was shown to be integrable if
\begin{equation}\label{pip:intcond}
\delta=-F^2G, \qquad K=FG.
\end{equation}%
By making use of the quasispin formalism \cite{talmi_simple_1993} and absorbing a phase in the quasispin operators, the Hamiltonian can be rewritten as
\begin{equation}\label{pip:ham}
\hat{H}=\delta\, b^{\dagger}b+\sum_{k=1}^m\epsilon_k S_k^0 -G\sum_{k,k'=1}^m\sqrt{\epsilon_k \epsilon_k'}S_k^{\dagger}S_{k'}-K\sum_{k=1}^m \sqrt{\epsilon_k}\left(S_k^{\dagger}b+S_kb^{\dagger}\right),
\end{equation}
which can again be related to a $su(2)$-based RG model. Starting from a hyperbolic Gaudin algebra \cite{ortiz_exactly-solvable_2005}
\begin{equation}
X_{ij}=2\frac{\sqrt{\epsilon_i \epsilon_k}}{\epsilon_i-\epsilon_k}, \qquad Z_{ij}=\frac{\epsilon_i+\epsilon_k}{\epsilon_i-\epsilon_k},
\end{equation}
for $(m+1)$ levels and exchanging a single $su(2)$ algebra with a pseudo-deformed $A(\xi)$-algebra, the Hamiltonian can be obtained as a linear combination of the conserved charges in the contraction limit. Labeling this single pseudo-deformed algebra as '$0$' and renormalizing the coupling constant and bosonic energy level $\epsilon_0$ as
\begin{equation}
g=\frac{\xi}{s_0+\xi \kappa}, \qquad \epsilon_0=\frac{\xi}{2s_0}\eta_0^2,
\end{equation}
the constants of motion become
\begin{align}\label{pip:number}
R_0^p &\equiv \lim_{\xi \to 0} R_0(\xi)=b^{\dagger}b+\sum_{k=1}^mS_k^0 \equiv N, \\
R_i^p &\equiv \lim_{\xi \to 0 } \frac{s_0}{\xi} R_i(\xi ) = \sum_{k \neq i}^m\left[\frac{\sqrt{\epsilon_i\epsilon_k}}{\epsilon_i-\epsilon_k}\left(S^{\dagger}_i S_k+S_i S^{\dagger}_k \right)+\frac{\epsilon_i+\epsilon_k}{\epsilon_i-\epsilon_k}S_i^0S_k^0\right] \nonumber\\
&\qquad \qquad \qquad \qquad +\frac{\eta_0}{\sqrt{\epsilon_i}}\left(S^{\dagger}_i b+S_i b^{\dagger}\right)+S_i^0 \left(\kappa+b^{\dagger}b-\frac{\eta_0^2}{\epsilon_i}\right),
\end{align}
with $N$ the number operator counting the number of excitations. Note that $R_i(\xi \to 0)=0$, but $R_i(\xi)/\xi$ remains finite for the whole range of $\xi$ and results in a non-zero conserved operator in the contraction limit. These operators are the building blocks for the Hamiltonian (\ref{pip:ham}), similar to the results presented for the fermionic $p_x+ip_y$ pairing model \cite{rombouts_quantum_2010}. We can take the linear combination
\begin{align}
\sum_{k=1}^m \epsilon_k R^p_k &= \sum_{k=1}^m\epsilon_k S_k^0\left(\kappa+b^{\dagger}b+\sum_{k' \neq k}^m S^0_{k'}\right)+\sum_{k=1}^m\sum_{k' \neq k}^m\sqrt{\epsilon_{k}\epsilon_{k'}}S^{\dagger}_kS_{k'} \nonumber\\
&\ \ +\eta_0 \sum_{k=1}^m \sqrt{\epsilon_k}\left(S_k^{\dagger}b+S_k b^{\dagger}\right)-\eta_0^2 \sum_{k=1}^m S_k^0,
\end{align}
then add the Casimir operators for each algebra $su(2)_k$ times $\epsilon_k$, and finally introduce the number operator (\ref{pip:number}). The resulting Hamiltonian now becomes exactly Eq. ({\ref{pip:ham})
\begin{align}
\hat{H}&=\eta_0^2 b^{\dagger}b+\sum_{k=1}^m (\kappa+N)\epsilon_k S_k^0+\sum_{k,k'=1}^m\sqrt{\epsilon_k \epsilon_{k'}}S^{\dagger}_k S_{k'} \nonumber \\
&\ \ +\eta_0  \sum_{k=1}^m\sqrt{\epsilon_k}\left(S_k^{\dagger}b+S_k b^{\dagger}\right)-\eta_0^2 N.
\end{align}
The integrability condition (\ref{pip:intcond}) arises naturally from the parametrization of the Gaudin algebra. The Hamiltonian studied in \cite{hibberd_bethe_2006} can be obtained by taking the linear combination
\begin{equation}
\hat{H}=\sum_{i=1}^m R_i^p = \sum_{i=1}^m \frac{\eta_0}{\sqrt{\epsilon_i}}\left(\Sd{i}b+S_i b^{\dagger}\right)-\sum_{i=1}^m \frac{\eta_0^2}{\epsilon_i}S_i^0 +\sum_{i=1}^m S_i^0(b^{\dagger}b+\kappa).
\end{equation}
The Bethe Ansatz states for these models can also be found from the contraction limit as
\begin{equation}
\ket{\psi}=\prod_{\alpha=1}^N\left(b^{\dagger}-\sum_{k=1}^m \frac{\sqrt{\epsilon_k}x_{\alpha}}{\epsilon_k-x_{\alpha}}\frac{S^{\dagger}_k}{\eta_0}\right)\ket{\theta},
\end{equation}
with resulting RG equations
\begin{equation}
\kappa-\frac{\eta_0^2}{x_{\alpha}}+\sum_{k=1}^m s_k \frac{\epsilon_k+x_{\alpha}}{\epsilon_k-x_{\alpha}}-\sum_{\beta \neq \alpha}^N\frac{x_{\beta}+x_{\alpha}}{x_{\beta}-x_{\alpha}}=0, \qquad \forall \alpha=1 \dots N.
\end{equation}
Using the techniques from \cite{claeys_eigenvalue-based_2015} (see Appendix \ref{app:ebv}), these can again be shown to be equivalent to a set of quadratic equations in the variables
\begin{align}\label{pip:deflam}
\Lambda_i&=\sum_{\alpha=1}^N\frac{\epsilon_i+x_{\alpha}}{\epsilon_i-x_{\alpha}}, \qquad \forall i=1 \dots m ,\\
\Lambda_0&=\sum_{\alpha=1}^N\frac{\eta_0^2}{x_{\alpha}},
\end{align}
which have to satisfy
\begin{align}\label{pip:eqlam}
\Lambda_i^2 &=-N(m-N)-2\kappa \Lambda_i +2 \eta_0^2\left(\Lambda_0+\frac{\Lambda_i+N}{\epsilon_i}\right)+\sum_{j \neq i}^m \frac{\epsilon_i+\epsilon_j}{\epsilon_i-\epsilon_j}(\Lambda_i-\Lambda_j), \qquad \forall i=1 \dots m, \\
2 \Lambda_0&=\sum_{i=1}^m \Lambda_i+2\kappa N.
\end{align}
The name 'eigenvalue-based variables' is apt since a single variable $\Lambda_i$ fully determines the eigenvalue of a single constant of motion $R_i^p$. If the rapidities satisfy the Richardson-Gaudin equations, we obtain
\begin{align}
R_i^p \ket{\psi} &= \frac{1}{2}\left[-\kappa-\sum_{\alpha=1}^N \frac{\epsilon_i+x_{\alpha}}{\epsilon_i-x_{\alpha}}+\frac{\eta_0^2}{\epsilon_i}+\sum_{k \neq i}^m \frac{\epsilon_i+\epsilon_k}{\epsilon_i-\epsilon_k}\right]\ket{\psi} \\
&=\frac{1}{2}\left[-\kappa-\Lambda_i+\frac{\eta_0^2}{\epsilon_i}+\sum_{k \neq i}^m \frac{\epsilon_i+\epsilon_k}{\epsilon_i-\epsilon_k}\right]\ket{\psi}.
\end{align}

\section{Investigation of the adiabatic connection}
%Eigenschappen van elke toestand - TDA oplossing - correspondentie sterke-zwakke koppelingslimiet - invloed s0 - kritische punten 
Since the pseudo-deformation provides an adiabatic connection between the bosonic algebra and the quasispin algebra, there is a similar connection between the Dicke model and an XXZ RG model. The adiabatic connection can be investigated in order to shed some light on the connection between the two models. When deforming a single level, the RG equations for arbitrary $\xi$ are given by 
\begin{equation}
1+ gs_0(\xi) Z_{0\alpha}+\frac{g}{2} \sum_{i=1}^m Z_{i\alpha} - g \sum_{\beta \neq \alpha}^N Z_{\beta \alpha}=0, \qquad \forall \alpha=1 \dots N.
\end{equation}
Inserting the Gaudin algebra  and the parametrization for the Dicke model in these equations, the RG equations at arbitrary values of $\xi$ are given by
\begin{equation}\label{ac:rgeq}
(\epsilon_0-E_{\alpha})-\sum_{i=1}^m s_i \frac{2G^2+\xi \epsilon_i E_{\alpha}/s_0}{\epsilon_i-E_{\alpha}}+ \sum_{\beta \neq \alpha}^N\frac{2G^2+\xi E_{\beta } E_{\alpha}/s_0}{E_{\beta}-E_{\alpha}}=0,\qquad \forall \alpha=1 \dots N.
\end{equation}
For $\xi=0$, these equations reduce to the equations for the Dicke model, while for $\xi=1$ we obtain RG equations for a trigonometric spin model
\begin{equation}\label{ac:rgeq1}
(\epsilon_0-E_{\alpha})-\sum_{i=1}^m s_i \frac{2G^2+\epsilon_i E_{\alpha}/s_0}{\epsilon_i-E_{\alpha}}+ \sum_{\beta \neq \alpha}^N\frac{2G^2+E_{\beta } E_{\alpha}/s_0}{E_{\beta}-E_{\alpha}}=0,\qquad \forall \alpha=1 \dots N.
\end{equation}
Interestingly these equations can also be seen to reduce to the regular Dicke model equations for an infinite $s_0$. This is easily understood as it corresponds to an infinite degeneracy and thus a bosonic mode, so the pseudo-deformation scheme would not change the physics of the problem. When solving equations (\ref{ac:rgeq}) for arbitrary $\xi$, it can be seen that singular points occur, similar to those occurring in the RG equations for $\xi=1$. In these singular points, multiple rapidities $x_{\alpha}$ coalesce with a single-particle level $\epsilon_i$, leading to diverging contributions to the RG equations. These divergencies cancel mutually exact, but have prevented straightforward numerical solutions of the RG equations for a long time \cite{rombouts_solving_2004,dominguez_solving_2006}. The occurrence of singular points can be linked to the Pauli exclusion principle \cite{de_baerdemacker_richardson-gaudin_2012}, and no singular points occur in a fully bosonic limit. However, since we are only dealing with a single bosonic mode coupled to spin modes, singular points will remain in these equations at every value of $\xi$. As an illustration, several solutions to the equations have been given in Figure \ref{fig:connect} for all values of the deformation parameter $\xi=0 \dots 1$, showing the qualitative behaviour of the solutions between both limits. As can be inferred from Figure \ref{fig:connect}, the qualitative behaviour of the connected eigenstates in the Dicke model and XXZ RG model can be quite different. For instance, the parameters used in the model are such that the Dicke model is in the weak-coupling limit (small $G$). Nevertheless, the connected XXZ RG state in the second column is clearly a collective state in the Cooper pairing regime \cite{dukelsky_colloquium:_2004}, as indicated by the complex-conjugate rapidities. The qualitative difference is accentuated by the presence of multiple singular points along the adiabatic path of the pseudo-deformation.

%In the regular RG equations, the coupling is fully determined by the single coupling-constant $G$, where the coupling between the equations disappears in the uncoupled limit $G=0$. We see here that in this limit the equations remain coupled through the second terms in the numerators, with an additional coupling determined by $\xi/s_0$. A consequence of this is that the weak-coupling limits of the Dicke model does not correspond to the weak-coupling limit of the trigonometric for small enough $s_0/\xi$. The correspondence between these models is illustrated in Figure \ref{fig:connect}.
\begin{figure}[ht]                      
 \begin{center}
 \includegraphics[width=\textwidth]{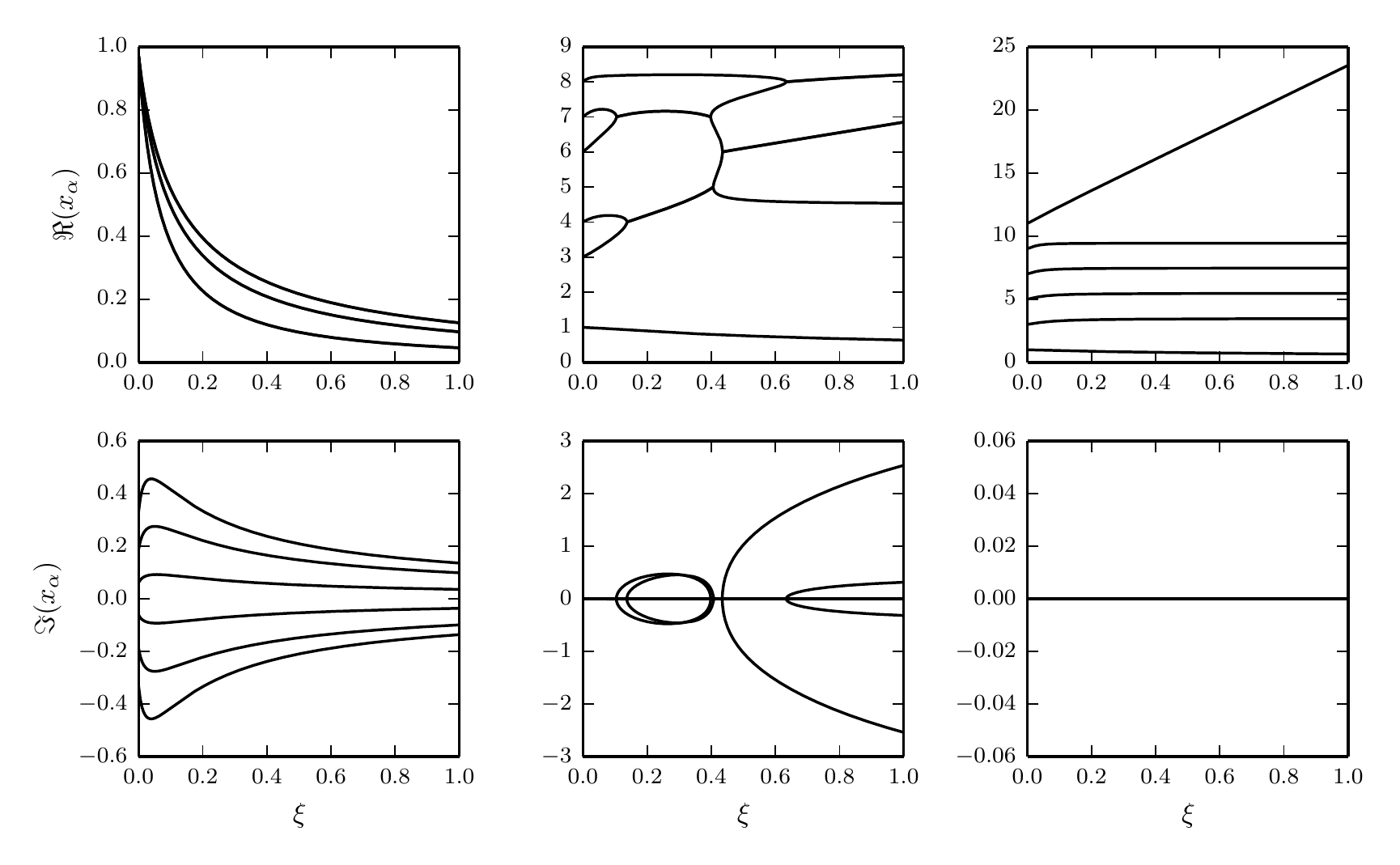}
 \caption{Several examples of the connection between the Dicke model and the XXZ model for $N=6$ excitations. Each column depicts the connection between different eigenstate of the models. The parameters are chosen such that the Dicke model (left side of each column) is in the weak-interaction regime ($G=-0.1$, $\epsilon_0=1$, $\{\epsilon_i\}=\{2,3,4, \dots, 12\}$ and $s_0=1/2$). Although the rapidities for the Dicke model approximately coincide with the single-particle levels, they do not end up in the weak-coupling limit for $\xi=1$ in the XXZ RG model. \label{fig:connect}}
 \end{center}
\end{figure}

\section{Deriving eigenvalue-based determinant expressions for form factors and overlaps}
%Probleem voor bosonische: geen echt pseudovacuum door oneindig-dimensionale representatie, Zelfde filosofie als daarjuist: oneindige limiet van RG model die gerenormalizeerd wordt door een juiste keuze van de variabelen
In the previous section, it was shown how many results for integrable systems containing a bosonic degree of freedom can be obtained in the contraction limit. However, if we wish to obtain expressions for normalizations and form factors starting from the RG models, we face the problem that the bosonic level has no state of maximum occupation number. A hole-vacuum now cannot be defined since this would contain an infinite number of bosonic excitations. This problem was first envisioned for the Dicke model by Tschirhart and Faribault \cite{tschirhart_algebraic_2014}, who devised an alternative formulation of the Algebraic Bethe Ansatz, introducing a pseudovacuum, which allowed for a description in terms of the eigenvalue-based variables for the Dicke model. As we will show, the pseudo-deformation scheme allows for a simpler derivation of these results by means of a renormalization procedure. Starting from the eigenvalue-based determinant expressions for the XXZ RG models \cite{claeys_eigenvalue-based_2015}, the dual state can be defined for any Bethe Ansatz state. By dividing the overlap between the Bethe Ansatz state and its dual state by the overlap between this dual state and a reference state (see section 6.2), a finite expression is obtained in the contraction limit, corresponding to the results previously presented \cite{tschirhart_algebraic_2014}. This approach can then be extended to the $(p+ip)$-wave pairing model.

\subsection{Overlap with non-interacting basis states}
%Tonen hoe grotere ontaardingen bekomen kunnen worden uit permanent door limiet te nemen van verschillende rapiditeiten die naar elkaar convergeren. eta0 terugroepen en tonen hoe dit de juiste uitdrukking levert. Borchardt ook eens vermelden maar niet expliciet geven.
A determinant expression for the overlap of an arbitrary Bethe Ansatz state in the Dicke model with a basis state (\ref{de:sd}) will first be derived. The expressions for these overlaps do not depend on the existence of a dual state, nor a hole-vacuum state, so the inclusion of a bosonic level into the existing fermionic $su(2)$ models in the pseudo-deformation is well defined. The main difference with the previously-considered spin-$1/2$ models is the occurrence of the bosonic level with arbitrary occupation number. The expansion (\ref{det:expansion}) with permanents as expansion coefficients (\ref{det:expansioncoeff}) still holds, but since multiple columns of the permanent can be equal, these can not immediately be rewritten as determinants. This problem can be avoided by introducing a limiting procedure. The permanent of a coefficient matrix with distinct columns can be rewritten as a determinant (\ref{ov:XXZ}), so we will introduce a matrix with different columns and consider the limit where multiple columns become equal. The overlaps can then be found from a two-step limiting procedure: first the permanent with multiple equal columns can be reduced to a determinant, after which the contraction limit can be taken.
%As will be demonstrated, the matrix representation of the coefficient matrix does not depend on the number of excitations of the degenerate level ($i=0$), so one can safely introduce the pseudo deformation for this level and take the contraction limit. 

Assume a model where all spins $s_i=1/2$, except for one level (again labeled $0$) with an arbitrary large degeneracy. The Bethe Ansatz can then be expanded in a set of basis states as
\begin{equation}
\prod_{\alpha=1}^N\left(X_{0\alpha}S_0^{\dagger}+\sum_{i=1}^mX_{i\alpha}\Sd{i}\right)\ket{\theta}=\sum_{[\{N_i\}]} \phi_{[\{N_i\}]} \left(\Sd{0}\right)^{N_0} \left(\prod_{a=1}^{N-N_0} \Sd{i_a} \right)\ket{\theta},
\end{equation}
with the vacuum state now containing the lowest-weight state $\ket{s_0,-s_0}$ for the degenerate level. The expansion coefficient is given by
\begin{equation}
\phi_{[\{N_i\}]}=\frac{1}{N_0!} \perm \left(
\begin{array}{cccccc}
X_{0\alpha_1} & \dots & X_{0\alpha_1} & X_{i_1 \alpha_1} & \dots & X_{i_{N-N_0} \alpha_1} \\
X_{0\alpha_2} & \dots & X_{0\alpha_2} & X_{i_1 \alpha_2} & \dots & X_{i_{N-N_0} \alpha_2} \\
\vdots &  & \vdots & \vdots &  & \vdots \\
\undermat{N_0}{X_{0\alpha_N} & \dots & X_{0\alpha_N}} & X_{i_1 \alpha_N} & \dots & X_{i_{N-N_0} \alpha_N} \\
\end{array}
\right).
\end{equation}
\vspace{\baselineskip}

\noindent 
In the following, it will be shown how this permanent can be written as the determinant of an $(N-N_0)\times(N-N_0)$-matrix for the trigonometric realization of the Gaudin algebra leading to the Dicke model. For this realization, the Gaudin algebra elements (\ref{dicke:param}) associated with the bosonic level are given by
\begin{equation}
X_{0\alpha}=\sqrt{1+\eta_{\alpha}^2}=\lim_{\eta_0 \to \infty}\frac{\sqrt{(1+\eta_0^2)(1+\eta_{\alpha}^2)}}{\eta_0-\eta_{\alpha}}
\end{equation}
Instead of immediately taking the limit $\eta_0 \to \infty$, it is possible to first evaluate the permanent for arbitrary $\eta_0$ and later take this limit. We now have a parametrization with a free parameter $\eta_0$ and wish to evaluate
\begin{equation}
\phi_{[\{N_i\}]}=\lim_{\eta_0 \to \infty}\frac{1}{N_0!}\perm \left(
\begin{array}{cccccc}
X_{0\alpha_1} & \dots & X_{0\alpha_1} & X_{i_1 \alpha_1} & \dots & X_{i_{N-N_0} \alpha_1} \\
X_{0\alpha_2} & \dots & X_{0\alpha_N} & X_{i_1 \alpha_2} & \dots & X_{i_{N-N_0} \alpha_2} \\
\vdots &  & \vdots & \vdots &  & \vdots \\
\undermat{N_0}{X_{0\alpha_N} & \dots & X_{0\alpha_N}} & X_{i_1 \alpha_N} & \dots & X_{i_{N-N_0} \alpha_N} \\
\end{array}
\right)
\end{equation}
\vspace{\baselineskip}

\noindent 
Instead of using the same parameter $\eta_0$ for each column and taking the limit for each column simultaneously, it is possible to introduce a different variable $\zeta_{i}$ (replacing $\eta_0$) for each column $i$, and taking consecutive limits to infinity of these parameters. This reduces the problem to the evaluation of
\begin{equation}
\phi_{[\{N_i\}]}=\frac{1}{N_0!}\lim_{\zeta_{1} \to \infty} \dots \lim_{\zeta_{{N_0}} \to \infty}\perm \left(
\begin{array}{cccccc}
X_{0_1\alpha_1} & \dots & X_{0_{N_0}\alpha_1} & X_{i_1 \alpha_1} & \dots & X_{i_{N-N_0} \alpha_1} \\
X_{0_1\alpha_2} & \dots & X_{0_{N_0}\alpha_2} & X_{i_1 \alpha_2} & \dots & X_{i_{N-N_0} \alpha_2} \\
\vdots &  & \vdots & \vdots &  & \vdots \\
\undermat{N_0}{X_{0_1\alpha_N} & \dots & X_{0_{N_0}\alpha_N}} & X_{i_1 \alpha_N} & \dots & X_{i_{N-N_0} \alpha_N} \\
\end{array}
\right)
\end{equation}
\vspace{\baselineskip}

\noindent 
with 
\begin{equation}
X_{0_i \alpha}=\frac{\sqrt{(1+\zeta_i^2)(1+\eta_{\alpha}^2)}}{\zeta_i-\eta_{\alpha}}.
\end{equation}
This is the permanent of a matrix where each matrix element satisfies the Gaudin algebra (\ref{rg:algebra}), for which a determinant representation exists
\begin{equation}
\phi_{[\{N_i\}]}=\frac{1}{N_0!}\lim_{\zeta_{1} \to \infty} \dots \lim_{\zeta_{{N_0}} \to \infty} \frac{\prod_{\alpha=1}^N\sqrt{1+\eta_{\alpha}^2}}{\prod_{a=1}^{N-N_0}\sqrt{1+\eta_{i_a}^2}\prod_{i=1}^{N_0}\sqrt{1+\zeta_{i}^2}} \det J,
\end{equation}
with $J$ defined as
\begin{equation}
 J_{ij} =
  \begin{cases}
   \eta_i+\sum_{\alpha}\frac{1+\eta_i \eta_{\alpha}}{\eta_i-\eta_{\alpha}}-\sum_{k \neq i}\frac{1+\eta_i\eta_k}{\eta_i-\eta_k} &\text{if}\ i=j \\
   \frac{\sqrt{1+\eta_i^2}\sqrt{1+\eta_j^2}}{\eta_i-\eta_j}       &\text{if}\ i \neq j
  \end{cases}.
\end{equation}
This is simply the application of Eq. (\ref{ov:XXZ}) with the auxiliary level tending to infinity. For the problem at hand, we can identify four different sectors in the matrix. If we associate the indices $i,j$ with levels tending to infinity and the indices $a,b$ with the finite levels, $J$ can be expressed as 
\begin{equation}
J=\left(\begin{array}{c|c}
J_{ij} & J_{ia}  \\ \hline
J_{ai} & J_{ab}
\end{array}\right),
\end{equation}
where the matrix elements of these sectors are given by
\begin{equation}
J_{ij}=\begin{cases}
\zeta_i+\sum_{\alpha=1}^N \frac{1+\zeta_i \eta_{\alpha}}{\zeta_i-\eta_{\alpha}}-\sum_{j\neq i}^{N_0}\frac{1+\zeta_i\zeta_j}{\zeta_i-\zeta_j}-\sum_{a=1}^{N-N_0} \frac{1+\zeta_i \eta_{a}}{\zeta_i-\eta_{a}} &\text{if}\ i=j \\
\frac{\sqrt{1+\zeta_i^2}\sqrt{1+\zeta_j^2}}{\zeta_i-\zeta_j} &\text{if}\ i\neq j
\end{cases},
\end{equation}
\begin{equation}
J_{ab}=\begin{cases}
\eta_a+\sum_{\alpha=1}^N \frac{1+\eta_a \eta_{\alpha}}{\zeta_i-\eta_{\alpha}}-\sum_{b\neq a}^{N-N_0}\frac{1+\eta_a\eta_b}{\eta_a-\eta_b}-\sum_{i=1}^{N} \frac{1+\eta_a \zeta_i}{\eta_a-\zeta_i} &\text{if}\ a=b \\
\frac{\sqrt{1+\eta_a^2}\sqrt{1+\eta_b^2}}{\eta_a-\eta_b} &\text{if}\ a\neq b
\end{cases},
\end{equation}
\begin{equation}
J_{ia}=-J_{ai}=\frac{\sqrt{1+\zeta_i^2}\sqrt{1+\eta_a^2}}{\zeta_i-\eta_a}.
\end{equation}
Absorbing the factors $(1+\zeta_{i}^2)^{1/4}\approx \sqrt{\zeta_{i}}$ from the prefactor into the first $N_0$ columns $i$ and the first $N_0$ rows $i$  and taking the subsequent limits to infinity as
\begin{equation}
\zeta_{1}\gg \zeta_{2}\gg \dots \gg \zeta_{N_0},
\end{equation}
the overlap can be written as
\begin{equation}
\phi_{[\{N_i\}]}=\frac{1}{N_0!} \frac{\prod_{\alpha=1}^N\sqrt{1+\eta_{\alpha}^2}}{\prod_{a=1}^{N-N_0}\sqrt{1+\eta_{i_a}^2}} \det J.
\end{equation}
Here all matrix elements were redefined as (with $\lim$ denoting the subsequent limits to infinity)
\begin{equation}
J_{ij}=\begin{cases}
\lim \frac{1}{\sqrt{1+\zeta_i^2}}\left[\zeta_i+\sum_{\alpha=1}^N \frac{1+\zeta_i \eta_{\alpha}}{\zeta_i-\eta_{\alpha}}-\sum_{k\neq i}^{N_0}\frac{1+\zeta_i\zeta_k}{\zeta_i-\zeta_k}-\sum_{a=1}^{N-N_0} \frac{1+\zeta_i \eta_{a}}{\zeta_i-\eta_{a}}\right] &\text{if}\ i=j \\
\lim \left[\frac{1}{\zeta_i-\zeta_j}(1+\zeta_i^2)^{1/4}(1+\zeta_j^2)^{1/4}\right] &\text{if}\ i \neq j
\end{cases},
\end{equation}
\begin{equation}
J_{ab}=\begin{cases}
\lim \left[\eta_a+\sum_{\alpha=1}^N \frac{1+\eta_a \eta_{\alpha}}{\eta_a-\eta_{\alpha}}-\sum_{b\neq a}^{N-N_0}\frac{1+\eta_a\eta_b}{\eta_a-\eta_b}-\sum_{i=1}^{N} \frac{1+\eta_a \zeta_i}{\eta_a-\zeta_i}\right] &\text{if}\ a=b \\
\lim \left[\frac{\sqrt{1+\eta_a^2}\sqrt{1+\eta_b^2}}{\eta_a-\eta_b}\right] &\text{if}\ a \neq b
\end{cases},
\end{equation}
\begin{equation}
J_{ia}=-J_{ai}=\lim \frac{1}{(1+\zeta_i^2)^{1/4}}\left[\frac{\sqrt{1+\zeta_i^2}\sqrt{1+\eta_a^2}}{\zeta_i-\eta_a}\right].
\end{equation}

All limits are straightforward except the limit of the diagonal elements in the first sector, which needs a few algebraic manipulations
\begin{align}
J_{ii}&=\lim \frac{1}{\sqrt{1+\zeta_i^2}}\left[\zeta_i+\sum_{\alpha=1}^N \frac{1+\zeta_i \eta_{\alpha}}{\zeta_i-\eta_{\alpha}}-\sum_{a=1}^{N-N_0} \frac{1+\zeta_i \eta_{a}}{\zeta_i-\eta_{a}}-\sum_{j\neq i}^{N_0}\frac{1+\zeta_i\zeta_j}{\zeta_i-\zeta_j}\right] \\
&=\lim  \frac{1}{\sqrt{1+\zeta_i^2}}\left[\zeta_i+\sum_{\alpha=1}^N\eta_{\alpha}-\sum_{a=1}^{N-N_0}\eta_{a} - \sum_{j \neq i}^{N_0}\frac{1+\zeta_i\zeta_j}{\zeta_i-\zeta_j}\right] \\
&=\lim  \frac{1}{\sqrt{1+\zeta_i^2}}\left[\zeta_i+\sum_{\alpha=1}^N\eta_{\alpha}-\sum_{a=1}^{N-N_0}\eta_{a} - \sum_{j <i}\frac{1+\zeta_i\zeta_j}{\zeta_i-\zeta_j}- \sum_{j >i}\frac{1+\zeta_i\zeta_j}{\zeta_i-\zeta_j}\right] \\
&=\lim  \frac{1}{\sqrt{1+\zeta_i^2}}\left[\zeta_i+\sum_{\alpha=1}^N\eta_{\alpha}-\sum_{a=1}^{N-N_0}\eta_{a} + \sum_{j <i}\zeta_i- \sum_{j >i} \zeta_j\right] \\
&=\lim \frac{1}{\sqrt{1+\zeta_i^2}}\left[i \zeta_i+\sum_{j>i} \zeta_j\right]=i. 
\end{align}
Now all matrix elements can be determined as
\begin{equation}\label{ov:Jab}
J_{ab}=\begin{cases}
(N_0+1)\eta_a+\sum_{\alpha=1}^N \frac{1+\eta_a \eta_{\alpha}}{\eta_a-\eta_{\alpha}}-\sum_{b\neq a}^{N-N_0}\frac{1+\eta_a\eta_b}{\eta_a-\eta_b} &\text{if}\ a=b \\
\frac{\sqrt{1+\eta_a^2}\sqrt{1+\eta_b^2}}{\eta_a-\eta_b} &\text{if}\ a \neq b,
\end{cases},
\end{equation}
\begin{equation}
J_{ij}=\begin{cases}
i &\text{if}\ i=j \\
0 &\text{if}\ i \neq j,
\end{cases},
\end{equation}
\begin{equation}
J_{ia}=-J_{ai}=0.
\end{equation}
In summary, the first $N_0$ diagonal elements become $1,2,\dots, N_0$, while the off-diagonal elements in the first $N_0$ rows reduce to zero. The expansion coefficient can then be rewritten as
\begin{eqnarray}
\phi_{[\{N_i\}]}&=&\frac{1}{N_0!}\frac{\prod_{\alpha}\sqrt{1+\eta_{\alpha}^2}}{\prod_{i \neq 0}\sqrt{1+\eta_{i}^2}} \det
\left[
\begin{array}{ccc|c}
1 & \dots & 0 & 0  \\
\vdots &  & \vdots &   \\
0 & \dots & N_0 & 0 \\\hline
0 & \dots &  & J_{ab} \\
\end{array}
\right]\nonumber\\
&=&\frac{\prod_{\alpha}\sqrt{1+\eta_{\alpha}^2}}{\prod_{i \neq 0}\sqrt{1+\eta_{i}^2}} \det [J],
\end{eqnarray}
with the $(N-N_0) \times (N-N_0)$ matrix $J_{ab}$ defined as in Eq. (\ref{ov:Jab}). This holds for arbitrary parameters $\{\eta_a,a=1 \dots N-N_0\}$ and $\{\eta_{\alpha}, \alpha=1 \dots N\}$. In order to obtain expressions for the Dicke model, it is possible to again introduce the parametrization (\ref{dicke:param}) and take the contraction limit. This immediately results in
\begin{equation}
\braket{N_0;\{i_a\}|\psi}=\sqrt{N_0!}(-G)^{N-N_0}\det[J]
\end{equation}
with
\begin{equation}
 J_{ab} =
  \begin{cases}
   \sum_{\alpha=1}^N\frac{1}{\epsilon_{i_a}-x_{\alpha}}-\sum_{c \neq a}^{N-N_0}\frac{1}{\epsilon_{i_a}-\epsilon_{i_c}} &\text{if}\ a=b \\
   \frac{1}{\epsilon_{i_a}-\epsilon_{i_b}}       & \text{if}\ a \neq b
  \end{cases}.
\end{equation}
A similar derivation for the $(p+ip)$-wave pairing Hamiltonian (\ref{pip:ham}) is given in Appendix \ref{app:pip}, where it is shown that the overlap of a Bethe Ansatz state
\begin{equation}
\ket{\psi}=\prod_{\alpha=1}^N\left(b^{\dagger}-\sum_i \frac{\sqrt{\epsilon_i}x_{\alpha}}{\epsilon_i-x_{\alpha}}\frac{S^{\dagger}_i}{\eta_0}\right)\ket{\theta},
\end{equation}
with a non-interacting state (\ref{de:sd}) is given by
\begin{equation}
\braket{N_0; \{i_a\}|\psi}=\frac{\sqrt{N_0!}}{\sqrt{\prod_a \epsilon_{i_a}}} \frac{\det J}{\eta_0^{N-N_0}}
\end{equation}
with $J$ defined as 
\begin{equation}
J_{ab}=\begin{cases}
\frac{1}{2}\sum_{\alpha=1}^N \frac{\epsilon_{a}+x_{\alpha}}{\epsilon_{i_a}-x_{\alpha}}-\sum_{c \neq a}^{N-N_0}\frac{1}{2}\frac{\epsilon_{i_a}+\epsilon_{i_c}}{\epsilon_{i_a}-\epsilon_{i_c}}-\frac{N_0+1}{2} &\text{if}\ a=b\\
\frac{\sqrt{\epsilon_{i_a}\epsilon_{i_b}}}{\epsilon_{i_a}-\epsilon_{i_b}}  &\text{if}\ a \neq b\\
\end{cases}.
\end{equation}
\subsection{Normalization}
%Overlap tussen duale en normale gedeeld door overlap duale met SD (=renormalizatie). Limieten nemen levert een eindige uitdrukking die overeenkomt met die van Tschirhart en Faribault
The procedure in the previous section can also be used to obtain the normalization of the Bethe Ansatz states for these models. Assuming a dual state $\ket{\psi'}$ exists for each eigenstate $\ket{\psi}$, which is identical but has a different normalization, we have
\begin{equation}
\ket{\psi'}=N_{\psi}\ket{\psi} \rightarrow \braket{\phi|\psi'}=N_{\psi}\braket{\phi|\psi},
\end{equation}
which holds for any arbitrary state $\ket{\phi}$, so that the norm of the eigenstate is given by
\begin{equation}
\braket{\psi|\psi}=\frac{\braket{\psi'|\psi}}{N_{\psi}}=\braket{\psi'|\psi}\frac{\braket{\phi|\psi}}{\braket{\phi|\psi'}},
\end{equation}
where we will choose the reference state $\ket{\phi}=\ket{N}\otimes\ket{\frac{1}{2},-\frac{1}{2}}_1\otimes \dots \otimes \ket{\frac{1}{2},-\frac{1}{2}}_m$ in order to keep the correspondence with the results obtained by Tschirhart and Faribault \cite{tschirhart_algebraic_2014} transparent. By making use of the pseudo-deformation, the degeneracy of the deformed level is given by $2s_0(\xi)+1$. The normalization will be calculated for an arbitrary degeneracy, after which the limit of an infinite degeneracy follows from the contraction limit:
\begin{align}
\braket{\psi'|\psi}&=\braket{\theta'|\prod_{\mu}\Sd{\mu}\prod_{\alpha}\Sd{\alpha}|\theta}=\frac{\prod_{\alpha}\sqrt{1+\eta_{\alpha}^2}\prod_{\mu}\sqrt{1+\eta_{\mu}^2}}{\prod_{i}\sqrt{1+\eta_i^2}} \det[J^{tot}]\braket{s_0(\xi)|(\Sd{})^{2s_0(\xi)}|-s_0(\xi)}, \\
\braket{\phi|\psi}&=\braket{\phi|\prod_{\alpha}\Sd{\alpha}|\theta}=\prod_{\alpha}\sqrt{1+\eta_{\alpha}^2} \braket{s_0(\xi)+N|(\Sd{})^N|-s_0(\xi)}, \\
\braket{\phi|\psi'}&=\braket{\phi|\prod_{\mu}S_{\mu}|\theta'}=\frac{\prod_{\mu}\sqrt{1+\eta_{\mu}^2}}{\prod_{i}\sqrt{1+\eta_i^2}} \det[J^{dual}] \braket{-s_0(\xi)+N|(S)^{2s_0(\xi)-N}|s_0(\xi)}.
\end{align}
The matrices are given by Eq. (\ref{ov:Jab}), with the number of pseudo-deformed creation/annihilation operators $N_0$ given by $2s_0(\xi)$, $N$ and $2s_0(\xi)-N$ respectively. Since not only the expansion coefficients are necessary but also the matrix elements, these have been written out explicitly.

Combining these expressions and evaluating the matrix elements, the result is
\begin{equation}
\braket{\psi|\psi}=\prod_{\alpha=1}^N(1+\eta_{\alpha}^2) \frac{(2s_0(\xi))!N!}{(2s_0(\xi)-N)!}\frac{\det[J^{tot}]}{\det[J^{dual}]},
\end{equation}
where the term $(2s_0(\xi))!N!/(2s_0(\xi)-N)!$ is obtained from the action of the generators on the irreps. Both  $J^{tot}$ and $J^{dual}$ are $m \times m$ matrices with matrix elements
\begin{equation}
 J^{tot}_{ij} =
  \begin{cases}
   2\sum_{\alpha=1}^N\frac{1+\eta_i \eta_{\alpha}}{\eta_i-\eta_{\alpha}}+\frac{2}{g}- \sum_{k \neq i,k\neq 0}^m\frac{1+\eta_i\eta_k}{\eta_i-\eta_k}+(2s_0(\xi)+1)\eta_i &\text{if}\ i=j \\
   \frac{\sqrt{1+\eta_i^2}\sqrt{1+\eta_j^2}}{\eta_i-\eta_j}       & \text{if}\ i \neq j
  \end{cases},
\end{equation}
\begin{equation}
 J^{dual}_{ij} =
  \begin{cases}
  \sum_{\alpha=1}^N\frac{1+\eta_i \eta_{\alpha}}{\eta_i-\eta_{\alpha}}+\frac{2}{g}- \sum_{k \neq i,k\neq 0}^m\frac{1+\eta_i\eta_k}{\eta_i-\eta_k}+(2s_0(\xi)-N+1)\eta_i &\text{if}\ i=j \\
   \frac{\sqrt{1+\eta_i^2}\sqrt{1+\eta_j^2}}{\eta_i-\eta_j}       & \text{if}\ i \neq j
  \end{cases}.
\end{equation}
Taking the contraction limit of these expressions and making use of the prefactor introduced in Eq. (\ref{dicke:ba}), a determinant expression for the Dicke model is obtained as
\begin{equation}
\braket{\psi|\psi}=N! \frac{\det[J^{tot}]}{\det[J^{dual}]}
\end{equation}
with
\begin{equation}
 J^{tot}_{ij} =
  \begin{cases}
   2\sum_{\alpha=1}^N\frac{1}{\epsilon_i-x_{\alpha}}+\frac{\epsilon_i-\epsilon_0}{G}- \sum_{k \neq i}^m\frac{1}{\epsilon_i-\epsilon_k} &\text{if}\ i=j \\
   \frac{1}{\epsilon_i-\epsilon_k}       & \text{if}\ i \neq j
  \end{cases},
\end{equation}
\begin{equation}
 J^{dual}_{ij} =
  \begin{cases}
   \sum_{\alpha=1}^N\frac{1}{\epsilon_i-x_{\alpha}}+\frac{\epsilon_i-\epsilon_0}{G}- \sum_{k \neq i}^m\frac{1}{\epsilon_i-\epsilon_k} &\text{if}\  i=j \\
   \frac{1}{\epsilon_i-\epsilon_k}       &\text{if}\ i \neq j
  \end{cases}.
\end{equation}
These correspond to the matrices derived previously \cite{tschirhart_algebraic_2014} and can be written in the set of eigenvalue-based variables
\begin{equation}
\Lambda_i=\sum_{\alpha=1}^N\frac{1}{\epsilon_i-x_{\alpha}},
\end{equation}
eliminating any explicit dependency on the rapidities $\{x_{\alpha}\}$.

Again, a similar derivation can be made for the extended $(p+ip)$-wave pairing model (see Appendix \ref{app:pip}), resulting in
\begin{equation}
\braket{\psi|\psi}=N! \frac{\det[J^{tot}]}{\det[J^{dual}]}
\end{equation}
\begin{equation}
 J^{tot}_{ij} =
  \begin{cases}
   \sum_{\alpha=1}^N\frac{\epsilon_i+x_{\alpha}}{\epsilon_i-x_{\alpha}}-\frac{\eta_0^2}{\epsilon_i}- \frac{1}{2}\sum_{k \neq i}^m\frac{\epsilon_i+\epsilon_j}{\epsilon_i-\epsilon_k}-\frac{1}{2} &\text{if}\ i=j \\
   \frac{\sqrt{\epsilon_i \epsilon_j}}{\epsilon_i-\epsilon_k}       &\text{if}\ i \neq j
  \end{cases},
\end{equation}
and a dual matrix with an additional factor $1/2$ in the first summation in the diagonal elements. Note the similarity between the diagonal elements of these matrices and the RG equations for both models. This similarity already arises at the level of the Gaudin algebra, where the RG equations are given by
\begin{equation}
1+gs_0(\xi)Z_{0\alpha}+\frac{g}{2}\sum_{i=1}^m Z_{i\alpha} - g \sum_{\beta \neq \alpha}^N Z_{\beta \alpha}=0,
\end{equation}
whereas the diagonal elements of the total matrix $(J^{tot}_{ii})$ are (up to the term $Z_{ri}$) given by
\begin{equation}
\frac{2}{g}+2\sum_{\alpha=1}^N Z_{i\alpha} - 2s_0(\xi)Z_{i0} - \sum_{j \neq i}^m Z_{ij} = \frac{2}{g}\left(1+gs_0(\xi)Z_{0i}+\frac{g}{2}\sum_{\alpha=1}^N Z_{i\alpha}-\frac{g}{2}\sum_{j\neq i}^m Z_{ij}\right),
\end{equation}
which resembles the RG equations, but with the roles of the energy levels and rapidities exchanged.

\subsection{Form factors}
Following the methods introduced in \cite{faribault_determinant_2012} and later used in \cite{tschirhart_algebraic_2014} and \cite{claeys_eigenvalue-based_2015}, determinant expressions for several form factors of local spin operators can be straightforwardly obtained by making use of the results presented in the previous sections. These form factors for the Dicke model were previously given by  \cite{tschirhart_algebraic_2014}, so we will restrict ourselves to form factors for the extended $(p+ip)$-wave pairing model. We only present final results, since the followed method is analogous to \cite{faribault_determinant_2012,tschirhart_algebraic_2014,claeys_eigenvalue-based_2015}.

\subsubsection{Local raising and lowering operators}
The expectation values of local raising and lowering operators $S^{\dagger}_i,S_i,b^{\dagger},b$ are non-zero between two states where the number of excitations differs by one. These can again be written by making use of the set of eigenvalue-based variables. For two states $\ket{\{x_{\alpha}\}}$ and $\ket{\{x_{\mu}\}}$ with $N$ and $N-1$ excitations respectively, we obtain
\begin{align}
\braket{\{x_{\alpha}\}|S_k^{\dagger}|\{x_{\mu}\}} = N! \frac{\det [J^k]}{\det [J^{dual}]}
\end{align}
with $J^k$ an $(m-1) \times (m-1)$ matrix with matrixelements
\begin{equation}
J^k_{ij}=\begin{cases}
\frac{1}{2}\left[\sum_{\alpha=1}^N\frac{\epsilon_i+x_{\alpha}}{\epsilon_i-x_{\alpha}} + \sum_{\mu=1}^{N-1}\frac{\epsilon_i+x_{\mu}}{\epsilon_i-x_{\mu}} -2\frac{\eta_0^2}{\epsilon_i}+2\kappa-\sum_{j \neq i, j \neq k}^m \frac{\epsilon_i+\epsilon_j}{\epsilon_i-\epsilon_j}-1\right]&\text{if}\ i=j \\
\frac{\sqrt{\epsilon_i \epsilon_k}}{\epsilon_i-\epsilon_k} &\text{if}\ i \neq j
\end{cases} \text{with}\ i,j \neq k,
\end{equation}
and $J^{dual}$ defined as
\begin{equation}
J^{dual}_{ij}=\begin{cases}
\frac{1}{2}\left[\sum_{\mu=1}^{N-1}\frac{\epsilon_i+x_{\mu}}{\epsilon_i-x_{\mu}} -2\frac{\eta_0^2}{\epsilon_i}+2\kappa-\sum_{j \neq i}^m \frac{\epsilon_i+\epsilon_j}{\epsilon_i-\epsilon_j}-1\right]&\text{if}\ i=j \\
\frac{\sqrt{\epsilon_i \epsilon_k}}{\epsilon_i-\epsilon_k} &\text{if}\ i \neq j
\end{cases}.
\end{equation}
For the bosonic creation operator, we have
\begin{align}
\braket{\{x_{\alpha}\}|b^{\dagger}|\{x_{\mu}\}} = N! \frac{\det [J^b]}{\det [J^{dual}]},
\end{align}
with the $m \times m$ matrix $J^b$ given by
\begin{equation}
J^b_{ij}=\begin{cases}
\frac{1}{2}\left[\sum_{\alpha=1}^N\frac{\epsilon_i+x_{\alpha}}{\epsilon_i-x_{\alpha}} + \sum_{\mu=1}^{N-1}\frac{\epsilon_i+x_{\mu}}{\epsilon_i-x_{\mu}} -2\frac{\eta_0^2}{\epsilon_i}+2\kappa-\sum_{j \neq i}^m \frac{\epsilon_i+\epsilon_j}{\epsilon_i-\epsilon_j}-1\right]&\text{if}\ i=j \\
\frac{\sqrt{\epsilon_i \epsilon_k}}{\epsilon_i-\epsilon_k} &\text{if}\ i \neq j
\end{cases}.
\end{equation}
The expressions for the annihilation operators can be found as the Hermitian conjugates of these operators.
\subsubsection{Local 'number' operators}
By making use of the Hellmann-Feynman theorem, the expectation value of local counting operators can be obtained. We will explicitly denote the eigenstates at a certain value of $\kappa$ as $\ket{x_{\alpha}(\kappa)}$ and $\ket{x_{\mu}(\kappa)}$ with eigenvalue-based variables $\{\Lambda_i^{\alpha}\}$ and $\{\Lambda_i^{\mu}\}$ respectively. From the definition of the constant of motion $R_i$, the expectation value of $S_i^0$ can be obtained as
\begin{equation}
\braket{x_{\alpha}(\kappa)|S_i^0|x_{\alpha}(\kappa)}=\frac{1}{2}\left(-1-\frac{\partial \Lambda_i^{\alpha}}{\partial \kappa}\right),
\end{equation}
with $\Lambda_i$ defined by Eq. (\ref{pip:deflam}), where the set of these variables have to satisfy Eq. (\ref{pip:eqlam}). By taking the partial derivative of these equations to $\kappa$, a linear system can be found for $\frac{\partial \Lambda_i}{\partial \kappa}$ as a function of $\Lambda_i$, similar to the method originally introduced in \cite{tschirhart_algebraic_2014}. Similarly, the off-diagonal expectation values can be found as
\begin{equation}
\braket{x_{\mu}(\kappa)|S_k^0|x_{\alpha}(\kappa)}=\frac{1}{2}\left(\Lambda_i^{\alpha}-\Lambda_i^{\mu}\right)\sum_{k=1}^m\frac{\partial \Lambda_k^{\alpha}}{\partial \kappa} \frac{\det \tilde{J}^k}{\det J^{dual}}
\end{equation}
with $J^{dual}$ defined as 
\begin{equation}
\tilde{J}^k_{ij}=\begin{cases}
\frac{1}{2}\left(\Lambda_i^{\alpha}+\Lambda_i^{\mu}-\sum_{l\neq i}^m\frac{\epsilon_i+\epsilon_l}{\epsilon_i-\epsilon_l}-\frac{2\eta_0^2}{\epsilon_i}+2\kappa-1\right) &\text{if}\ i= j\\
\frac{\sqrt{\epsilon_- \epsilon_j}}{\epsilon_i-\epsilon_j} &\text{if}\ i\neq j
\end{cases}.
\end{equation}
Matrix elements for $b^{\dagger}b$ can then be found by noting that $N=b^{\dagger}b+\sum_{i=1}^n S_i^0$, and the expectation values of the counting operator are known.

\section{Conclusions}
%Algemene methode om eigenvalue-based resultaten te bekomen voor RG integreerbare modellen met een bosonische vrijheidsgraad. Numeriek efficient, laat toe om decoherentie en zo te bestuderen.

In the present paper we have shown how two different classes of integrable models can be obtained within one unifying framework, using the pseudo-deformation of a single quasispin. Starting from a trigonometric and a hyperbolic realization of the XXZ models, we were able to derive the integrability of the Dicke model and the extended $(p+ip)$-pairing model, respectively. Furthermore, this connection was then used to link the determinant expressions for overlaps, normalizations and form factors with the determinant expressions recently presented by us for XXZ RG models \cite{claeys_eigenvalue-based_2015}. We were able to rederive the results for the Dicke model \cite{tschirhart_algebraic_2014}, circumventing the need for introducing the Algebraic Bethe Ansatz and the pseudovacuum, and generalize them toward the second class of integrable models containing a bosonic degree of freedom. Whereas strongly correlated systems in general exhibit unfavourable numerical scaling, expressions such as the ones presented in this paper allow for extensive numerical investigations of integrable systems. These results also further extend the description of RG integrable systems in terms of a new set of eigenvalue-based variables, explicitly symmetric in the rapidities, to models containing a bosonic degree of freedom. 

\section*{Acknowledgements}
Pieter W. Claeys received a Ph.D. fellowship and Stijn De Baerdemacker a postdoctoral fellowship from the Research Foundation Flanders (FWO Vlaanderen).

\appendix
\section{Equations for the eigenvalue-based variables}
\label{app:ebv}
For an XXZ model defined by a Gaudin algebra, it was previously shown \cite{claeys_eigenvalue-based_2015} how the set of RG equations (\ref{rg:rgeq}) are equivalent to a set of equations for the variables
\begin{equation}
\Lambda_i=\sum_{\alpha=1}^N Z_{i\alpha},
\end{equation}
where we considered a system with $n$ levels labelled $i$  and $N$ excitations labelled $\alpha$. These equations are given by
\begin{equation}\label{app:eqlam}
\Lambda_i^2=\Gamma N (1-N+2\sum_{j \neq i}^n s_j)-\frac{2}{g}\Lambda_i+2\sum_{j \neq i}^n s_j Z_{ij}\left(\Lambda_i-\Lambda_j\right) + (1-2s_i)\sum_{\alpha=1}^N Z_{i\alpha}^2,
\end{equation}
where $\Gamma$ is a constant defined by the Gaudin algebra as
\begin{equation}
X_{ij}^2-Z_{ij}^2=\Gamma, \qquad \forall i \neq j.
\end{equation}
It can be seen that these equations are closed in the set of variables $\{\Lambda_i\}$ if $s_i=1/2, \forall i$. Starting from an $(m+1)$-level system and introducing the pseudo-deformation of a single level $0$ and taking $s_i=1/2, \forall i \neq 0$, a closed set of equations can similarly be obtained. The equations for the set of $\Lambda_i$ are then given by
\begin{equation}
\Lambda_i^2=\Gamma N (m-N + 2 \frac{s_0}{\xi}) - \frac{2}{g}\Lambda_i + \sum_{j \neq i, j \neq 0}^m Z_{ij}(\Lambda_i-\Lambda_j)+2\frac{s_0}{\xi}Z_{i0}(\Lambda_i-\Lambda_0).
\end{equation}
By introducing the proposed parametrizations for both models models, a closed set of equations can be obtained in the contraction limit.

\subsection{Dicke model}
Starting from the trigonometric representation ($\Gamma=1$) for the Dicke model, we have
\begin{equation}
\Lambda_0=\sum_{\alpha=1}^N \eta_{\alpha}, \qquad \Lambda_i=\sum_{\alpha=1}^N \frac{1+\eta_i\eta_{\alpha}}{\eta_i-\eta_{\alpha}}, \qquad i=1 \dots m,
\end{equation}
which have to satisfy
\begin{align}
\left[\sum_{\alpha=1}^N \frac{1+\eta_i\eta_{\alpha}}{\eta_i-\eta_{\alpha}}\right]^2=&N(m-N+2s_0/\xi)-\frac{2}{g}\sum_{\alpha=1}^N \frac{1+\eta_i \eta_{\alpha}}{\eta_i-\eta_{\alpha}} \nonumber\\
&+\sum_{j \neq i}^m \frac{1+\eta_i \eta_j}{\eta_i-\eta_j}\left(\sum_{\alpha=1}^N \frac{1+\eta_i\eta_{\alpha}}{\eta_i-\eta_{\alpha}}-\frac{1+\eta_j\eta_{\alpha}}{\eta_j-\eta_{\alpha}}\right) - 2\frac{s_0}{\xi}\eta_i \left(\sum_{\alpha=1}^N \frac{1+\eta_i\eta_{\alpha}}{\eta_i-\eta_{\alpha}}+\eta_{\alpha}\right).
\end{align}
Introducing the proposed expressions for $\eta_i$ and $g$, multiplying everything with  $2s_0/\xi$, and taking the contraction limit immediately simplifies this to a set of equations
\begin{equation}
G^2 [\Lambda_i^D]^2=N-\Lambda_i^D (\epsilon_i-\epsilon_0)+G^2 \sum_{j \neq i}^m \frac{\Lambda_i^D-\Lambda_j^D}{\epsilon_i-\epsilon_j},
\end{equation}
in the variables
\begin{equation}
\Lambda_i^D=\sum_{\alpha=1}^N \frac{1}{\epsilon_i-x_{\alpha}}.
\end{equation}

\subsection{$(p+ip)$-wave pairing}
Starting from the hyperbolic parametrization ($\Gamma=-1$), we have
\begin{equation}
\Lambda_0=\sum_{\alpha=1}^N \frac{\xi \eta_0^2 + 2 s_0 x_{\alpha}}{\xi \eta_0^2-2s_0x_{\alpha}}, \qquad \Lambda_i=\sum_{\alpha=1}^N \frac{\epsilon_i+x_{\alpha}}{\epsilon_i-x_{\alpha}},
\end{equation}
where the following expansion holds for $\xi \ll 1$
\begin{equation}
\Lambda_0=\sum_{\alpha=1}^N \left(-1-\frac{\xi}{s_0}\frac{\eta_0^2}{x_{\alpha}}\right)+\mathcal{O}(\xi^2)= -N-\frac{\xi}{s_0}\sum_{\alpha=1}^N \frac{\eta_0^2}{x_{\alpha}}+\mathcal{O}(\xi^2),
\end{equation}
where we can define $\Lambda_0^p=\sum_{\alpha=1}^N \eta_0^2/x_{\alpha}$. Introducing this parametrization in the set of equations, we obtain
\begin{equation}
\Lambda_i^2=-N(m-N+\frac{2s_0}{\xi})-2\left(\frac{s_0}{\xi}+\kappa\right)\Lambda_i+\sum_{j \neq i}^m \frac{\epsilon_i+\epsilon_j}{\epsilon_i-\epsilon_j}(\Lambda_i-\Lambda_j)+\frac{2s_0}{\xi}\left(\frac{2s_0\epsilon_i+\xi \eta_0^2}{2s_0\epsilon_i-\xi \eta_0^2}\right)(\Lambda_i+N+\frac{\xi}{s_0}\Lambda_0^p)+\mathcal{O}(\xi),
\end{equation}
where all terms containing $1/\xi$ drop out. In the contraction limit, this results in
\begin{equation}
\Lambda_i^2=-N(m-N)-2\kappa \Lambda_i + \sum_{j \neq i}^n \frac{\epsilon_i+\epsilon_i}{\epsilon_i-\epsilon_j}(\Lambda_i-\Lambda_j)+\frac{2\eta_0^2}{\epsilon_i}(\Lambda_i+N)+2\Lambda_0^p.
\end{equation}
An additional equation can be found linking $\Lambda_0^p$ to the set of $\Lambda_i$ by considering (\ref{app:eqlam}) for the pseudo-deformed level. Keeping only the terms to dominant order in $\xi$, we obtain
\begin{equation}
N^2=-N(1-N+m)-2\left(\frac{2s_0}{\xi}+\kappa\right)(-N-\frac{\xi}{s_0}\Lambda_0^p)+\sum_{j=1}^m(N+\Lambda_j)+\left(1-\frac{2s_0}{\xi}\right)\left(N+\frac{2\xi}{s_0}+\Lambda_0^p\right)+\mathcal{O}(\xi),
\end{equation}
where all terms containing $1/\xi$ again drop out. In the contraction limit we are left with
\begin{equation}
-2\Lambda_0^p+\sum_{j=1}^m \Lambda_j = 2 \kappa N,
\end{equation}
where we have obtained a closed set of equations in the eigenvalue-based variables.

\section{Results for the $(p+ip)$-wave pairing model}
\label{app:pip}
In this Appendix, it is shown how overlaps for the extended $(p+ip)$-wave pairing model can be derived in the contraction limit of a hyperbolic RG model. The Bethe Ansatz state is given by
\begin{align}
\ket{\psi}&=\lim_{\xi \to 0}\prod_{\alpha=1}^N\left(X_{0\alpha}A^{\dagger}(\xi)+\sum_{i=1}^m X_{i\alpha}\Sd{i}\right)\ket{\theta} \nonumber\\
&=\prod_{\alpha=1}^N\left(\frac{-2 b^{\dagger}}{\sqrt{x_{\alpha}}}+2\sum_{i=1}^m \frac{\sqrt{\epsilon_i x_{\alpha}}}{\epsilon_i-x_{\alpha}}\Sd{i}\right)\ket{\theta} \nonumber\\
&=\sum_{[\{N_i\}]} \phi_{[\{N_i\}]} \left(b^{\dagger}\right)^{N_0}\prod_{i=1}^{N-N_0} \left(\Sd{i}\right)^{N_i}\ket{\theta}
\end{align}
for the parametrization proposed in Section \ref{pip}. The expansion coefficient for a state $\ket{N_0;\{i_a\}}$ is given by
\begin{equation}
\phi_{[\{N_i\}]}=\lim_{\xi \to 0}\frac{1}{N_0!}\left(\frac{2s_0}{\xi}\right)^{\frac{N_0}{2}}\perm \left(
\begin{array}{cccccc}
X_{0\alpha_1} & \dots & X_{0\alpha_1} & X_{i_1 \alpha_1} & \dots & X_{i_{N-N_0} \alpha_1} \\
X_{0\alpha_2} & \dots & X_{0\alpha_2} & X_{i_1 \alpha_2} & \dots & X_{i_{N-N_0} \alpha_2} \\
\vdots &  & \vdots & \vdots &  & \vdots \\
\undermat{N_0}{X_{0\alpha_N} & \dots & X_{0\alpha_N}} & X_{i_1 \alpha_N} & \dots & X_{i_{N-N_0} \alpha_N} \\
\end{array}
\right),
\end{equation}
\vspace{\baselineskip}

\noindent 
with $X_{ij}=2\sqrt{\epsilon_i\epsilon_j}/(\epsilon_i-\epsilon_j)$ and $\epsilon_0=\xi \eta_0^2/(2s_0)$. Introducing a different deformation parameter for each column, and relabelling $\epsilon_0=\xi/(2s_0 \eta_0^2)$ as $\zeta_i$ in column $i$, this permanent can be rewritten as 
\begin{equation}
\phi_{[\{N_i\}]}=\frac{1}{\eta_0^{N_0}N_0!}\frac{\sqrt{\prod_{\alpha=1}^Nx_{\alpha}}}{\sqrt{\prod_{a=1}^{N-N_0}\epsilon_a}} \lim_{\xi_1 \to 0} \dots \lim_{\xi_{N_0} \to 0}\det J,
\end{equation}
with the $N \times N$ matrix $J$ again consisting of four blocks, defined as
\begin{equation}
J=\left(\begin{array}{c|c}
J_{ij} & J_{ia}  \\ \hline
J_{ai} & J_{ab}
\end{array}\right),
\end{equation}
with
\begin{equation}
J_{ij}=\begin{cases}
\sum_{\alpha=1}^N \frac{\zeta_i+x_{\alpha}}{\zeta_i-x_{\alpha}}-\sum_{k\neq i}^{N_0}\frac{\zeta_i+\zeta_k}{\zeta_i-\zeta_k}-\sum_{a=1}^{N-N_0} \frac{\zeta_i +\epsilon_{a}}{\zeta_i-\epsilon_{a}} -1 &\text{if}\ i=j\\
2\frac{\sqrt{\zeta_i \zeta_j}}{\zeta_i-\zeta_j} &\text{if}\ i \neq j
\end{cases},
\end{equation}
\begin{equation}
J_{ab}=\begin{cases}
\sum_{\alpha=1}^N \frac{\epsilon_a+x_{\alpha}}{\epsilon_a-x_{\alpha}}-\sum_{c\neq a}^{N-N_0}\frac{\epsilon_a+\epsilon_c}{\epsilon_a-\epsilon_c}-\sum_{i=1}^{N_0} \frac{\epsilon_a+ \zeta_i}{\epsilon_a-\zeta_i}-1 &\text{if}\ a=b\\
\frac{2\sqrt{\epsilon_a\epsilon_b}}{\epsilon_a-\epsilon_b} &\text{if}\ a \neq b
\end{cases},
\end{equation}
\begin{equation}
J_{ia}=-J_{ai}=\frac{\sqrt{\zeta_i\epsilon_a}}{\zeta_i-\epsilon_a},
\end{equation}
where we have used the notation $\epsilon_{i_a} \equiv \epsilon_a$. Taking the limits to zero as
\begin{equation}
\zeta_1 \ll \zeta_2 \ll \dots \ll \zeta_{N_0},
\end{equation}
this results in
\begin{align}
J_{ii}&=-N-\sum_{j<i}\frac{\zeta_i+\zeta_j}{\zeta_i-\zeta_j}-\sum_{j>i}\frac{\zeta_i+\zeta_j}{\zeta_i-\zeta_j}+(N-N_0) -1\\
&=-N-(i-1)+(N_0-i)+(N-N_0)-1 =-2i \\
J_{ij}&=0 \\
J_{aa}&=\sum_{\alpha=1}^N \frac{\epsilon_a+x_{\alpha}}{\epsilon_a-x_{\alpha}}-\sum_{b\neq a}^{N-N_0}\frac{\epsilon_a+\epsilon_b}{\epsilon_a-\epsilon_b}-N_0-1\\
J_{ab}&=\frac{2\sqrt{\epsilon_a\epsilon_b}}{\epsilon_a-\epsilon_b} \\
J_{ia}&=-J_{ai}=0.
\end{align}
The expansion coefficient is then given by
\begin{align}
\phi_{[\{N_i\}]}&=\frac{1}{\eta_0^{N_0}N_0!}\frac{\sqrt{\prod_{\alpha=1}^Nx_{\alpha}}}{\sqrt{\prod_{a=1}^{N-N_0}\epsilon_a}} \det 
\left(\begin{array}{ccc|c}
-2 & \dots & 0 & 0  \\
\vdots &  & \vdots &   \\
0 & \dots & -2N_0 & 0 \\\hline
0 & \dots & 0 & J_{ab} \\
\end{array}\right) \nonumber\\
&=\frac{(-2)^{N_0}}{\eta_0^{N_0}}\frac{\sqrt{\prod_{\alpha=1}^Nx_{\alpha}}}{\sqrt{\prod_{a=1}^{N-N_0}\epsilon_a}} \det J_{ab},
\end{align}
which implies that the overlap of the wavefunction
\begin{equation}
\ket{\psi}=\prod_{\alpha=1}^N \left(b^{\dagger}-\sum_{i=1}^N \frac{\sqrt{\epsilon_i}x_{\alpha}}{\epsilon_i-x_{\alpha}}\frac{\Sd{i}}{\eta_0}\right)\ket{\theta}
\end{equation}
with a non-interacting state $\ket{\phi}=\ket{N_0;\{i_a\}}$ is given by
\begin{equation}
\braket{\phi|\psi}=\frac{\sqrt{N_0!}\sqrt{\prod_{a=1}^{N-N_0}\epsilon_a}}{(-\eta_0)^{N-N_0}} \det J,
\end{equation}
with  $J$ an $(N-N_0) \times (N-N_0)$ matrix defined as
\begin{equation}
J_{ab}=
\begin{cases}
\frac{1}{2}\sum_{\alpha=1}^N \frac{\epsilon_a+x_{\alpha}}{\epsilon_a-x_{\alpha}}-\frac{1}{2}\sum_{c \neq a}^{N-N_0}\frac{\epsilon_a+\epsilon_c}{\epsilon_a-\epsilon_c}-\frac{N_0+1}{2}
\qquad &\text{if}\ a=b\\
\frac{\sqrt{\epsilon_a\epsilon_b}}{\epsilon_a-\epsilon_b} \qquad &\text{if}\ a \neq b
\end{cases}.
\end{equation}
The expression for the normalization can immediately be taken from the Dicke model, so we have
\begin{equation}
\braket{\psi|\psi}=N! \frac{\det [J^{tot}]}{\det [J^{dual}]}.
\end{equation}
The off-diagonal elements of both matrices are given by $\sqrt{\epsilon_i \epsilon_j}/(\epsilon_i-\epsilon_j)$, while the diagonal elements of the top matrix are given by
\begin{align}
2J^{tot}_{ii}&=\lim_{\xi \to 0}2 \sum_{\alpha=1}^N \frac{\epsilon_i+x_{\alpha}}{\epsilon_i-x_{\alpha}}+\frac{2}{g}-\sum_{j \neq i}^n \frac{\epsilon_i+\epsilon_j}{\epsilon_i-\epsilon_j}-\sum_{j=1}^{2s_0(\xi)}\frac{\epsilon_i+\epsilon_0}{\epsilon_i-\epsilon_0}-1 \nonumber\\
&=\lim_{\xi \to 0}2 \sum_{\alpha=1}^N \frac{\epsilon_i+x_{\alpha}}{\epsilon_i-x_{\alpha}}+2\left(\frac{s_0}{\xi}+\kappa\right)-\sum_{j \neq i}^n \frac{\epsilon_i+\epsilon_j}{\epsilon_i-\epsilon_j}-\frac{2s_0}{\xi}\frac{\epsilon_i+\xi \eta_0^2/2s_0}{\epsilon_i-\xi \eta_0^2/2s_0}-1 \nonumber \\
&=\lim_{\xi \to 0}2 \sum_{\alpha=1}^N \frac{\epsilon_i+x_{\alpha}}{\epsilon_i-x_{\alpha}}+\frac{2s_0}{\xi}+2\kappa-\sum_{j \neq i}^n \frac{\epsilon_i+\epsilon_j}{\epsilon_i-\epsilon_j}-\frac{2s_0}{\xi}\left(1+\frac{\xi \eta_0^2}{s_0 \epsilon_i}\right)-1\nonumber \\
&=2 \sum_{\alpha=1}^N \frac{\epsilon_i+x_{\alpha}}{\epsilon_i-x_{\alpha}}-\sum_{j \neq i}^n \frac{\epsilon_i+\epsilon_j}{\epsilon_i-\epsilon_j}-\frac{2\eta_0^2}{\epsilon_i}+2\kappa-1,
\end{align}
and the diagonal elements of the bottom matrix are similarly found as
\begin{equation}
2J^{dual}_{ii}=\sum_{\alpha=1}^N \frac{\epsilon_i+x_{\alpha}}{\epsilon_i-x_{\alpha}}-\sum_{j \neq i}^n \frac{\epsilon_i+\epsilon_j}{\epsilon_i-\epsilon_j}-\frac{2\eta_0^2}{\epsilon_i}+2\kappa-1
=2J^{dual}_{ii}-\sum_{\alpha=1}^N\frac{\epsilon_i+x_{\alpha}}{\epsilon_i-x_{\alpha}}.
\end{equation}
\bibliographystyle{ieeetr}
\bibliography{MyLibrary}
\end{document}